\documentclass{aastex631}

  \usepackage{amsmath}
  \usepackage{graphicx}
  \usepackage[utf8]{inputenc}
  \graphicspath{{./}{figures/}}

  \received{--}
  \revised{--}
  \accepted{--}
  
  \shorttitle{Solar differential rotation}
\begin{document}

\title{Helioseismic Properties of Dynamo Waves in the Variation of Solar Differential Rotation}

   \author[0000-0003-3067-288X]{Krishnendu Mandal}
  \affiliation{New Jersey Institute of Technology, Newark, NJ 07102, USA}
  
  \author[0000-0003-0364-4883]{Alexander G. Kosovichev}
  \affiliation{New Jersey Institute of Technology, Newark, NJ 07102, USA}
  \affiliation{NASA Ames Research Center, Moffett Field, CA 94035, USA}
  \author[0000-0001-9884-1147]{Valery V. Pipin}
  \affiliation{Institute of Solar-Terrestrial Physics, Russian Academy of Sciences, Irkutsk, 664033, Russia}
    \begin{abstract}
   Solar differential rotation exhibits a prominent feature: its cyclic variations over the solar cycle, referred to as zonal flows or torsional oscillations, are observed throughout the convection zone. Given the challenge of measuring magnetic fields in subsurface layers, understanding deep torsional oscillations becomes pivotal in deciphering the underlying solar dynamo mechanism. In this study, we address the critical question of identifying specific signatures within helioseismic frequency-splitting data associated with the torsional oscillations. To achieve this, a comprehensive forward modeling approach is employed to simulate the helioseismic data for a dynamo model that, to some extent, reproduces solar-cycle variations of magnetic fields and flows. We provide a comprehensive derivation of the forward modeling process utilizing generalized spherical harmonics, as it involves intricate algebraic computations. All estimated frequency-splitting coefficients from the model display an $11$-year periodicity. Using the simulated splitting coefficients and realistic noise, we show that it is possible to identify the dynamo wave signal present in the solar zonal flow from the tachocline to the solar surface. By analyzing observed data, we find similar dynamo wave patterns in the observational data from MDI, HMI, and GONG. This validates the earlier detection of dynamo waves and holds potential implications for the solar dynamo theory models. 
        
  \end{abstract}
  
 \section{Introduction} \label{sec:intro}
  \citet{parker_55_dynamo} showed that dynamo action involves cyclic transformations of the poloidal and toroidal magnetic fields of the Sun. This scenario suggests that the magnetic field of bipolar magnetic regions is formed from the large-scale toroidal magnetic field that is generated from the axisymmetric poloidal magnetic field, twisted by differential rotation deep in the convection zone. As the solar dynamo revolves around comprehending the cyclic evolution of toroidal and poloidal magnetic fields, there is a consensus on the generation of the toroidal field. However, the generation process of the poloidal field remains less understood. \citet{parker_55_dynamo} suggested that the poloidal magnetic field is generated from the toroidal field by electromotive force excited by cyclonic convection which is known as ``$\alpha-$effect".  Several alternative mechanisms for the poloidal field generation can be found in the literature \citep{babcock61,arnab_dikpati99,paul11}. An alternative scenario, known as the flux transport model proposed by \citet{babcock61,leighton69}, posits that the cyclical evolution of solar activity is influenced by meridional circulation. This circulation is responsible for transporting the magnetic field of decaying sunspots from low latitudes toward the polar region, as outlined by \citet{dikpati09}. The transported magnetic field subsequently undergoes reconnection with the polar field of the preceding solar cycle and migrates to the base of the convection zone. In the tachocline, this magnetic field is carried equatorward by the meridional flow at the base of the convection zone. Eventually, it surfaces due to buoyancy instability, resulting in the formation of observable sunspots. This scenario gains validation through the observed migration of sunspots, originating from mid-latitudes at the onset of the solar cycle and progressively migrating towards the equator by its culmination. Observational evidence further indicates that the polar magnetic field attains its peak during sunspot minima, with its magnitude correlating with the subsequent cycle's sunspot number \citep{schatten78}. This information enables forecasting the magnitude of the upcoming solar cycle \citep{arnab07}. Using this technique, the forecast window spans $5$ to $6$ years, half the duration of the $11$-year solar cycle. 
  This supports the propositions by \citet{parker_55_dynamo} and \citet{babcock61} that the toroidal magnetic field is generated through the stretching of the poloidal magnetic field in polar regions. Whether this conveyor belt scenario, which depends on the structure of the meridional flow, is responsible for flux transport dynamo was under scrutiny. Helioseismology has found that it can either be single cell \citep{rajaguru15,mandal18,gizon20_mer} or double cell meridional flow \citep{zhao13,chen17}. However, all the studies to detect the deep structure of the meridional flow suffer from centre-to-limb systematics in the time-distance helioseismology measurements \citep{zhao12}. The contribution from the systematic and random errors is comparable in magnitude to the signal itself \citep{Stejko2022}. Aside from this difficulty, both the surface flux transport and dynamo-waves can reproduce the basic observed features of magnetic field evolution observed on the solar surface, including the correlation of the polar magnetic field during the magnetic cycle minimum with the subsequent sunspot cycle magnitude. The surface observations are not sufficient to distinguish between two dynamo scenarios. Currently, we can not measure sub-surface magnetic fields. Thus, the information about the dynamo process comes from the measurement of large-scale subsurface flows such as differential rotation and meridional flow. It is expected that variations of these flows over the scale of $11$ years are due to the solar-cycle evolution of subsurface magnetic fields. It is to be noted that these variations are not only due to magnetic torque. Other inertial forces can also contribute to this variation.  Nevertheless observed variations in subsurface flows provide an important clue about the mechanism of solar dynamo. 
 \par
 The torsional oscillations were discovered by \citet{howard80}. After subtracting the mean rotation, the data revealed alternating zones of fast and slow zonal flow bands, which originated at mid-latitudes and subsequently migrated to the equator just like the magnetic butterfly diagram. It was then suggested that these zone flows (called torsional oscillations because of their cyclic variations) are due to the action of the magnetic field of active regions. The observed zonal flow pattern is long-living and coherent over essentially the whole solar circumference. Therefore, it can not be of convective origin but may be induced by the dynamo-generated magnetic fields and associated processes. Numerical 3D simulation by \citet{gustavo16} suggested that torsional oscillation in the model is due to the magnetic torque at the base of the convection zone. However, modulation of the turbulent heat transport by the magnetic field in the convection zone results in variations in the zonal and meridional flows, producing the `extended' 22-year cycle of the torsional oscillations \citep{pipin2019}.
 Helioseismology can probe the subsurface flow profile of solar zonal flow \citep{howe18,basu19}.  \citet{sasha97} found that torsional oscillations are extended beneath the solar surface. Later \citet{vorontsov02} found that torsional oscillations are present in the entire convection zone with the phase propagating poleward and equatorward from mid-latitude at all depths throughout the convective envelope. \citet{pipin2019} suggested that torsional oscillation can be helpful for advanced prediction of solar cycles, $1-2$ sunspot cycles ahead. \citet{sasha2019} argued that the amplitude of zonal acceleration in the high-latitude region of the base of the convection zone during solar maxima may give information about the strength of following solar cycle maxima. If confirmed, this relationship would give the solar activity forecast a full $11$-year cycle ahead. \citet{pipin20} argued that the sign of the correlation between the flow amplitude in the deep convection zone and the strength of the subsequent solar cycle depends on the properties of long-term variations of the dynamo cycles. 
 
 Recognizing the significance of conclusions drawn from the aforementioned studies, we systematically revisit the problem. We conduct synthetic tests to validate the detection of the dynamo wave patterns within the solar zonal flow. Utilizing the variations of the zonal flow velocity from the dynamo model proposed by \citet{pipin20}, we calculate associated coefficients of rotational splitting of frequencies of solar oscillations corresponding to the p-modes observed by the MDI. HMI, and GONG instruments. The flow variations depicted in Figure \ref{fig:dynamo_model}, feature two distinct branches of faster-than-average rotational components akin to the observed solar zonal flows.  Our aim is to determine the precision with which we can reconstruct the zonal flows in the solar convection zones, from the tachocline to the solar surface by using helioseismic data. Through these tests, we demonstrate the high accuracy achievable in capturing the zonal flow across this entire depth range if the observational data are averaged over two-year periods. We then implement this methodology for measuring the zonal flow across the convection zone using the current helioseismic data from MDI, HMI, and GONG.
 
 \section{Data analysis and methods}
 \subsection{Rotational Frequency Splitting Data}
  We utilize a comprehensive set of global helioseismology data products obtained from various instruments, including space-based ones like the Helioseismic Magnetic Imager (HMI) \citep[HMI;][]{hmi} on board the Solar Dynamic Observatory (SDO) and Michelson Doppler Imager (MDI) \citep[MDI;][]{scherrer95} on board the Solar and Heliospheric Observatory (SoHO), as well as ground-based instrument Global Oscillation Network Group (GONG) \citep[GONG;][]{GONG}. Each acoustic mode is characterized by a triplet of quantum numbers, denoted as $(n,\ell,m)$, where $n$ signifies the radial order, $\ell$ represents the spherical harmonic degree, and $m$ designates the azimuthal order. In a spherically symmetric model, the mode frequency $\nu_{n \ell m}$ remains independent of the azimuthal order, $m$. However, the presence of axisymmetric perturbations, such as differential rotation, breaks this degeneracy in $m$ which can be expressed as 
  \begin{equation}
      \nu_{n\ell m}=\nu_{n\ell}+\sum_{s}^{s_{\text{max}}} a_{s}(n,\ell)\mathcal{P}_{s}^{\ell}(m),
      \label{eq:nu_lmn}
\end{equation}
  where $\nu_{n\ell}$ is the central frequency of a frequency multiple of radial order $n$ and angular degree $\ell$; $a_s$ are the rotational splitting coefficients (called the a-coefficients).  The odd-order coefficients, $a_1, a_3, a_5$, etc., capture information about the solar rotation. $\mathcal{P}_{s}^{\ell}$ are polynomials of degree $s$ \citep{ritzwoller91}, and these polynomials have orthogonal properties. The MDI data encompass the period from May $1996$ to February $2011$, while the SDO/HMI data used in this study cover the period from April $2010$ to May $2023$. We collect solar frequency data and their corresponding splitting information from distinct, non-overlapping $72$-day time series. The time series data for the MDI and HMI instruments are sourced from the SDO Joint Science Operations Center (JSOC) website\footnote{JSOC: \url{http://jsoc.stanford.edu/}}.  To investigate the temporal evolution of solar zonal flow across the entire duration of our analysis, we utilize a consolidated dataset comprising both MDI and HMI instruments. In the overlapping time frame of both instruments, we rely on the data exclusively from HMI. 
 \par
 GONG's data span from May $1995$ to January $2023$. The data are categorized by GONG's ``months", each spanning a duration of 36 days. Solar oscillation frequencies and splitting coefficients are derived through the analysis of a 108-day period, which is equal to observation of 3 GONG's months. There is an overlap of $72-$ days between different data sets.  
 
  \begin{figure}     
      \includegraphics[scale=0.5]{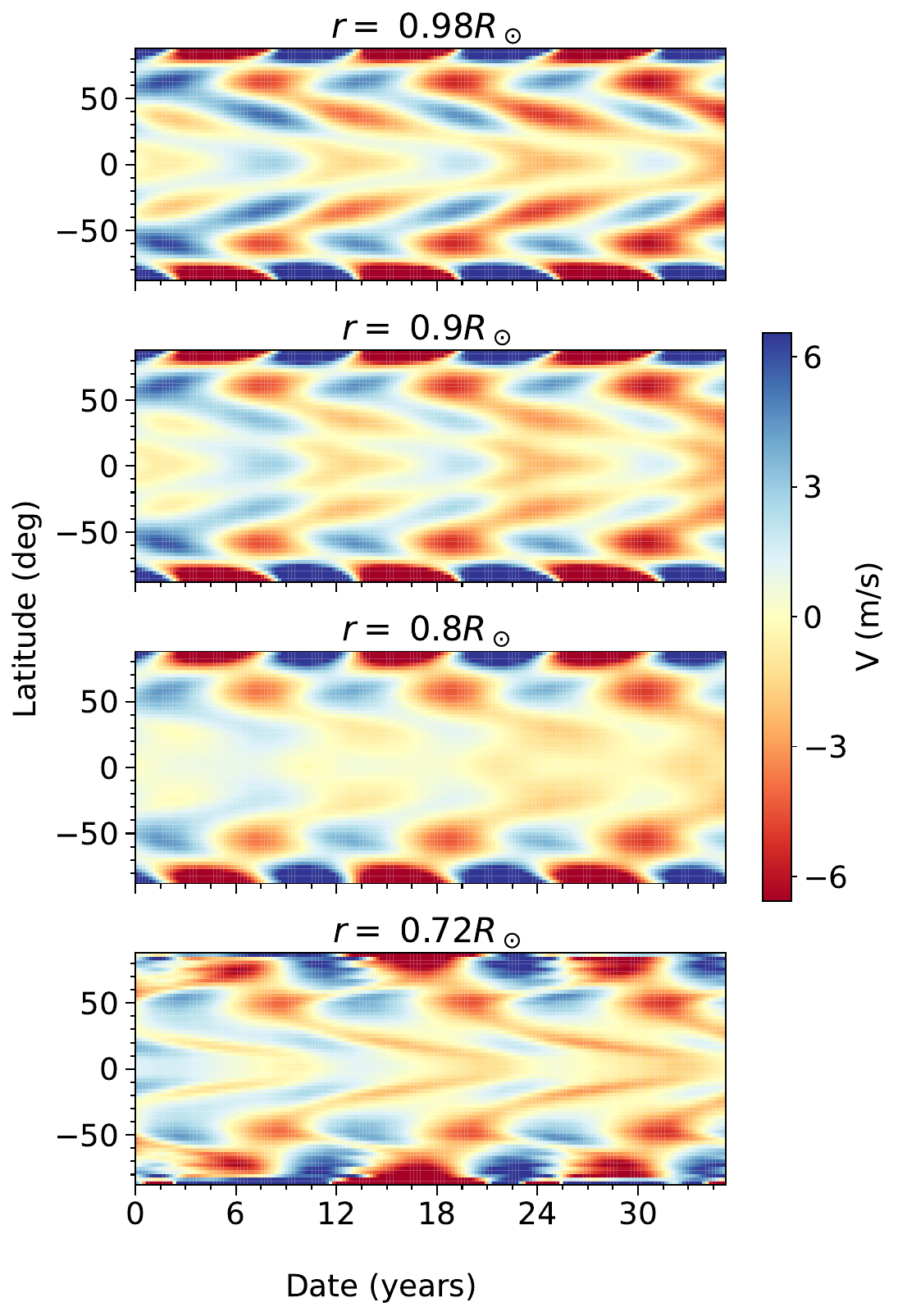}\includegraphics[scale=0.5]{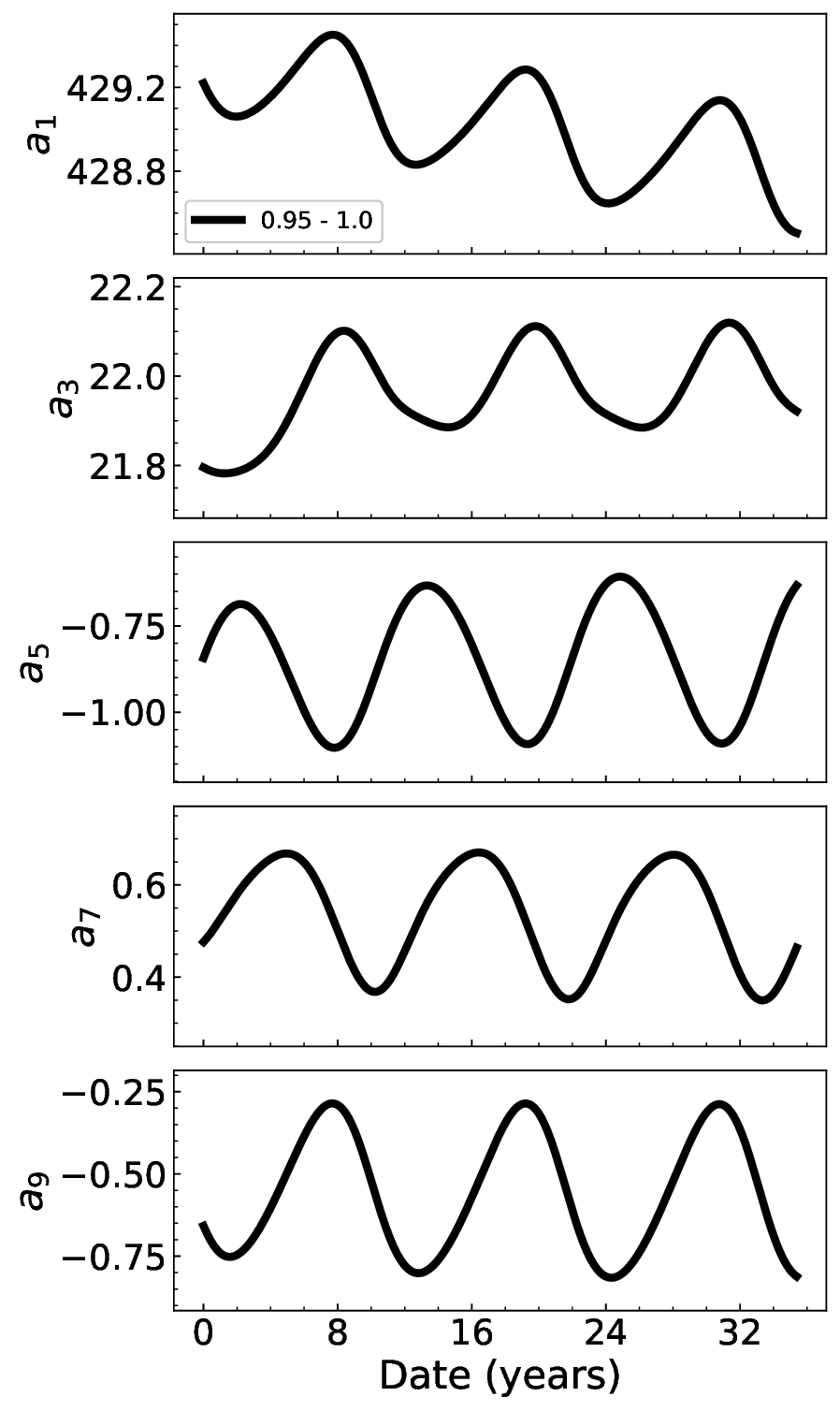}
      \caption{The zonal flow velocities, derived from the dynamo model of \citet{pipin20}, are depicted across various depths. Notably, these velocities exhibit both equatorward and poleward migrating branches stemming from the mid-latitude area. Employing this zonal flow velocity, forward modeling is conducted to estimate splitting coefficients. The right panel illustrates the temporal evolution of splitting coefficients measured in nHz.}
      \label{fig:dynamo_model}
  \end{figure}
  
  \subsection{Forward Modelling Approach}
   The solar rotational velocity can be expressed as the following 
   \begin{equation}
    v(r,\theta)=\sum_{s=1,3,5, \cdots} w_{s}(r)\partial_{\theta} Y_{s,0}(\theta,\phi)
    \label{eq:v_w}
   \end{equation}
   where $Y_{s,0}(\theta,\phi)$ is a spherical harmonic of angular degree $s$ and zero azimuthal order,  $w_s(r)$ are the radial coefficients in the velocity expansion, $r$ is the radius from the solar center, $\theta$ is the co-latitude, and $\phi$ is the azimuthal angle.
   Rotation leads to the splitting of p and f-mode frequencies with respect to the azimuthal order, $m$, of the oscillation eigenfunctions. This phenomenon can be expressed through the following integral equation: 
   \begin{equation}
       a_{s}(n,\ell)=\int_{0}^{R} K_{s}(n,\ell;r)w_{s}(r)r^2 dr,
       \label{eq:a_coeff}
   \end{equation}
   Here, $K_{s}(n,\ell)$ represents the sensitivity kernel for the mode $(n,\ell),$ determining the extent to which this specific mode is influenced by rotation, characterized by a spatial scale $s$. The precise expression of this function can be found in the appendix (see Appendix \ref{sec:app_diffRot}). By leveraging vector spherical harmonics, we have formulated the kernel expression. A similar approach was previously employed by \citet{ritzwoller91}. The beauty of using the generalized spherical harmonics lies in their capability to enable expressing derivatives with respect to $\theta$ and $\phi$ using vector spherical harmonics itself. Consequently, the entire derivation transforms into an elegant algebraic exercise. The use of generalized spherical harmonics has recently become prevalent in studies focusing on normal mode coupling \citep[see e.g.][]{hanasoge17_etal}. For an in-depth derivation, please refer to Section \ref{sec:app_diffRot}. We present the entire derivation due to the availability of the splitting a-coefficients using various normalizations. This often necessitates converting these coefficients into a preferred form for ease of use.
   Equation \ref{eq:a_coeff} allows us to estimate the a-coefficients provided we have a known rotation profile, denoted as $v(r,\theta)$, which is decomposed into a spherical harmonic basis to derive $w_s$ using Equation \ref{eq:v_w}.

   \begin{figure}     
     \includegraphics[scale=0.45]{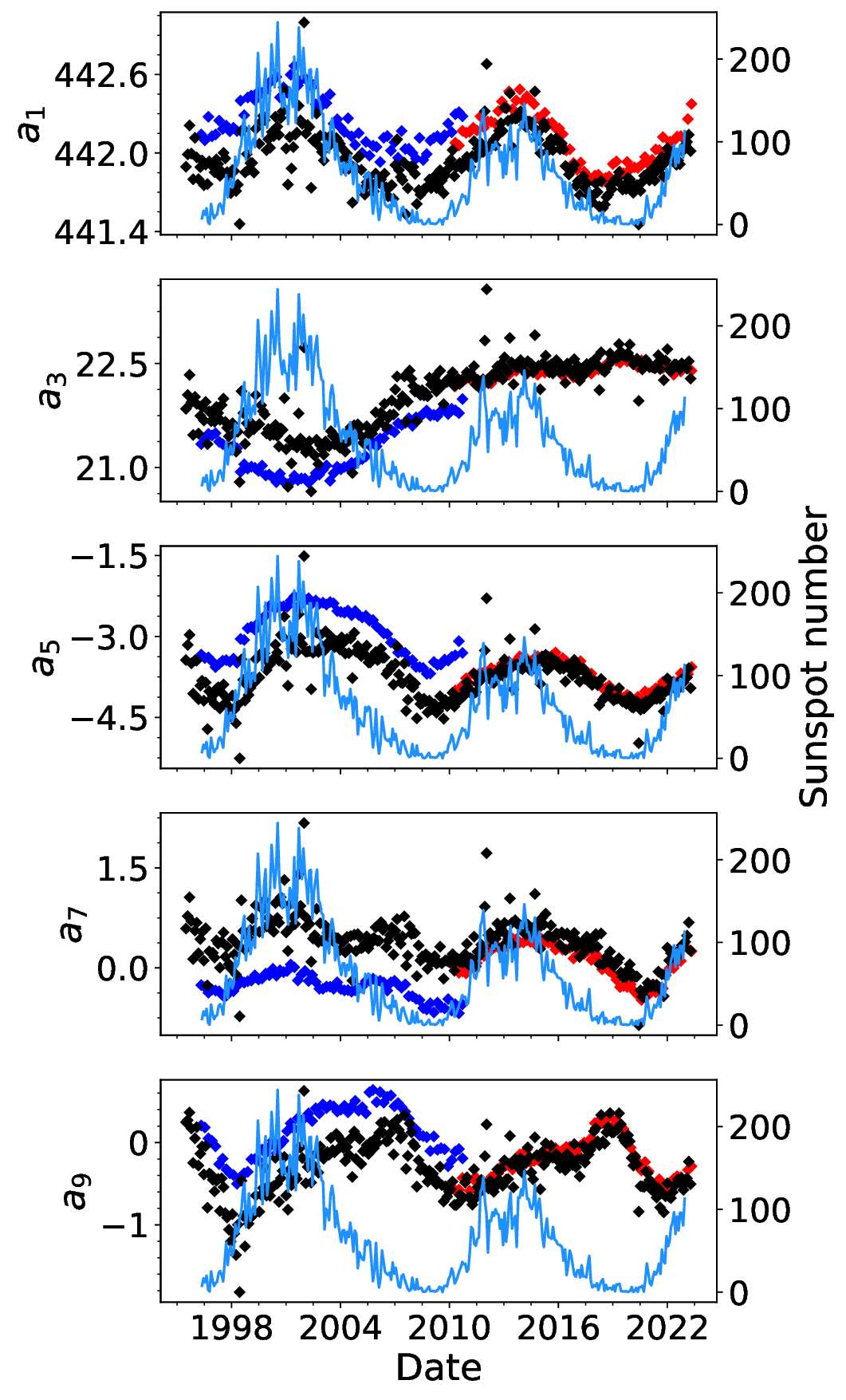}\hspace{1cm}\includegraphics[scale=0.45]{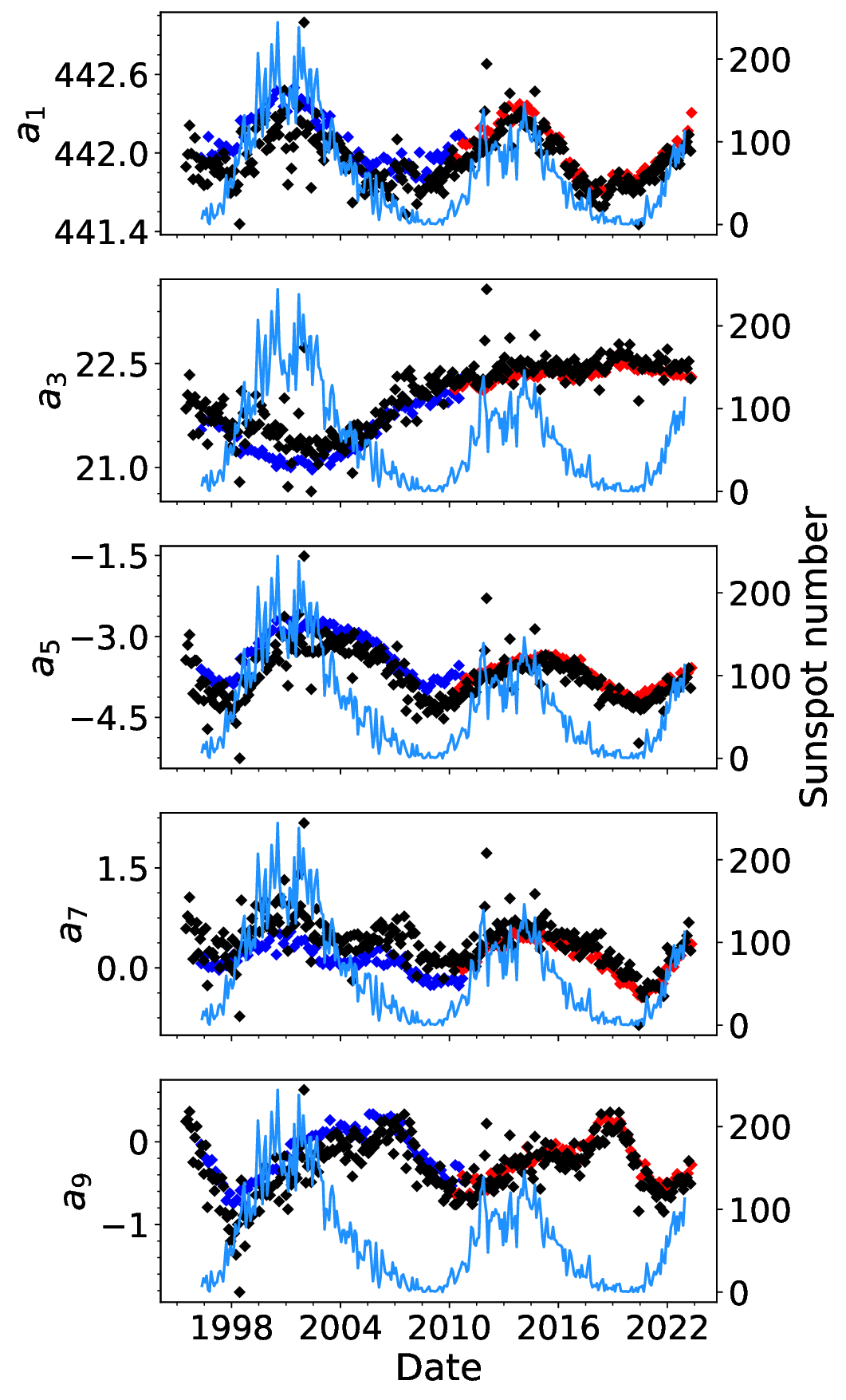}     
     \caption{The splitting coefficients' variation over time is depicted here. Left panel: we compare average a-coefficients measured in nHz for all the modes with the turning point radius of the p-modes between $0.9 R_\odot$ and $1R_\odot$ by using data from three instruments, SOHO/MDI (blue marker), SDO/HMI (red marker), and GONG (black marker). Right panel: we limit our analysis to a maximum harmonic degree of $120$ for MDI. This results in consistency across all three instruments.}
     \label{fig:a_coef_obs}
 \end{figure}
 
 \subsection{Inversion Method}
   In the preceding section, we discussed the application of forward modeling using a predefined rotation or zonal flow profile. These artificially generated a-coefficients serve as the basis for inversion, enabling an assessment of our ability to accurately replicate the zonal flow. To enhance the realism of the inversion process, random noise is introduced based on observed data. Employing a regularized least squares formula, we aim to minimize the misfit function, denoted as $\chi$, as the following
 \begin{equation}
   \chi^2=\sum_{n,\ell}\frac{1}{\sigma_{n,\ell}^{2}}\left(a_{s}(n,\ell)-\int K_{n,\ell,s}(r)w_{s}(r)r^{2}dr\right)^{2}+\lambda\int\left(\frac{\partial^{2}w_{s}}{\partial r^{2}}\right)^{2}r^{2}dr,
\label{eq:chi}
\end{equation}
 We expect our solution for $w_s(r)$ to manifest as a linear combination of splitting coefficients $a_s$.
 \begin{align}
     w_s(r_0)=\sum_{i\in M}c_{i}(r_0) a_{s}(i)\nonumber\\
     =\sum_{i\in M} c_i(r_0) \int K_{i,s}(r)w_{s}(r)r^{2}dr\nonumber\\
     =\int \mathcal{K}(r;r_0)w_s(r)r^2dr.
     \label{eq:avg_kernel_main}
 \end{align}
 Equation \ref{eq:avg_kernel_main} reveals that our inverted $w_s$ at depth $r_0$ represents a convolution of the true $w_s(r)$ profile, with $\mathcal{K}$ representing the averaging kernel, $M$ represents the set of measurements for all accessible acoustic modes. In Section \ref{sec:avg_kernel_app}, we derive the expression for the averaging kernel $\mathcal{K}(r;r_0)$ and demonstrate its behavior at various depths. Figure \ref{fig:avg_kernel} illustrates that the averaging kernel tends to be narrower near the surface and widens with greater depths. This observation underscores our higher resolution capability close to the surface compared to the base of the convection zone.
 To attain a smooth solution for $w_s(r)$, we employ second-order derivative smoothing. The function, $w_s$, is expanded on the basis of B-spline functions with $50$ equally spaced knots along the acoustic depth, ranging from the surface to the center. Additional details regarding our inversion method are available in Appendix \ref{sec:app_inv}. To compute error bars, we utilize the Monte Carlo method. This involves perturbing the measured splitting coefficients in accordance with observed noise, generating a sample set of $100$ realizations. Subsequently, we conduct inversions with each sample to derive profiles for $w_s(r)$. The variation among these different profiles of $w_s(r)$ allows us to calculate the standard deviation, which in turn quantifies the error in $w_s$. We then use Equation \ref{eq:v_w} to estimate the error in velocity, $v(r,\theta)$. Our inversion technique follows a $1.5$D approach. Other inversion methods are also available; for instance, refer to \citet{schou94,schou98}.

  \section{Results}
\subsection{Forward Modeling of Helioseismic Effects of Torsional Oscillations} 
 In their research, \citet{pipin20} delved into the connection between the amplitude of the ``extended" mode of migrating zonal flows and solar magnetic cycles using a nonlinear mean-field solar dynamo model. Their findings unveiled a notable correlation between torsional oscillation parameters, such as zonal acceleration, and the subsequent cycle magnitudes, albeit with a time lag ranging from $11$ to $20$ years. This discovery holds great significance, suggesting that helioseismic observations of torsional oscillations can offer valuable insights for more advanced predictions of solar cycles, potentially forecasting solar activity $1$ to $2$ sunspot cycles in advance. Given the substantial importance of this finding, we employ zonal flow data from their simulation and engage in a forward modeling approach. This allows us to gain insight into the specific signatures we should be seeking within observed a-coefficients. This result is depicted in Figure \ref{fig:dynamo_model}. We consider modes with turning-point radii between 0.95-1R$_{\odot}$. We calculate the average values of the a-coefficients of those mode sets. We only compare the variations of the a-coefficients with time. Upon examining Figure \ref{fig:dynamo_model}, it becomes evident that periodicity is present among all the a-coefficients across all harmonic degrees. To determine if similar observations can be made in the actual observed data, we conduct identical tests. We focused on the modes with the turning-point radii ranging from 0.9-1R$_\odot$. The results for the observed a-coefficients from all three instruments are shown in Figure \ref{fig:a_coef_obs}. Similar to the findings of \citet{basu19}, we see periodicity in all the a-coefficients in the observed data except for $a_3$, which does not exhibit periodicity during the available time period of data. It is possible that it reflects  a longer trend. While the splitting coefficients from GONG and HMI exhibit consistency (refer to the left panel of Figure \ref{fig:a_coef_obs}), those from MDI do not align. Following the approach of \citet{basu19,antia08}, we limit our analysis to a maximum harmonic degree of $120$ for MDI, which results in consistency across all three instruments (see right panel of Figure \ref{fig:a_coef_obs}). 
 To unveil the signature of dynamo waves from the tachocline to the solar surface, we must address the presence of outliers in our observed data. Given the relatively small magnitude of the signal, we can either eliminate outliers through data cleaning or extend the duration of the time series to enhance the signal-to-noise ratio.     
  
   \subsection{Inversion for synthetic tests}   
\begin{figure}     
     \includegraphics[scale=0.5]{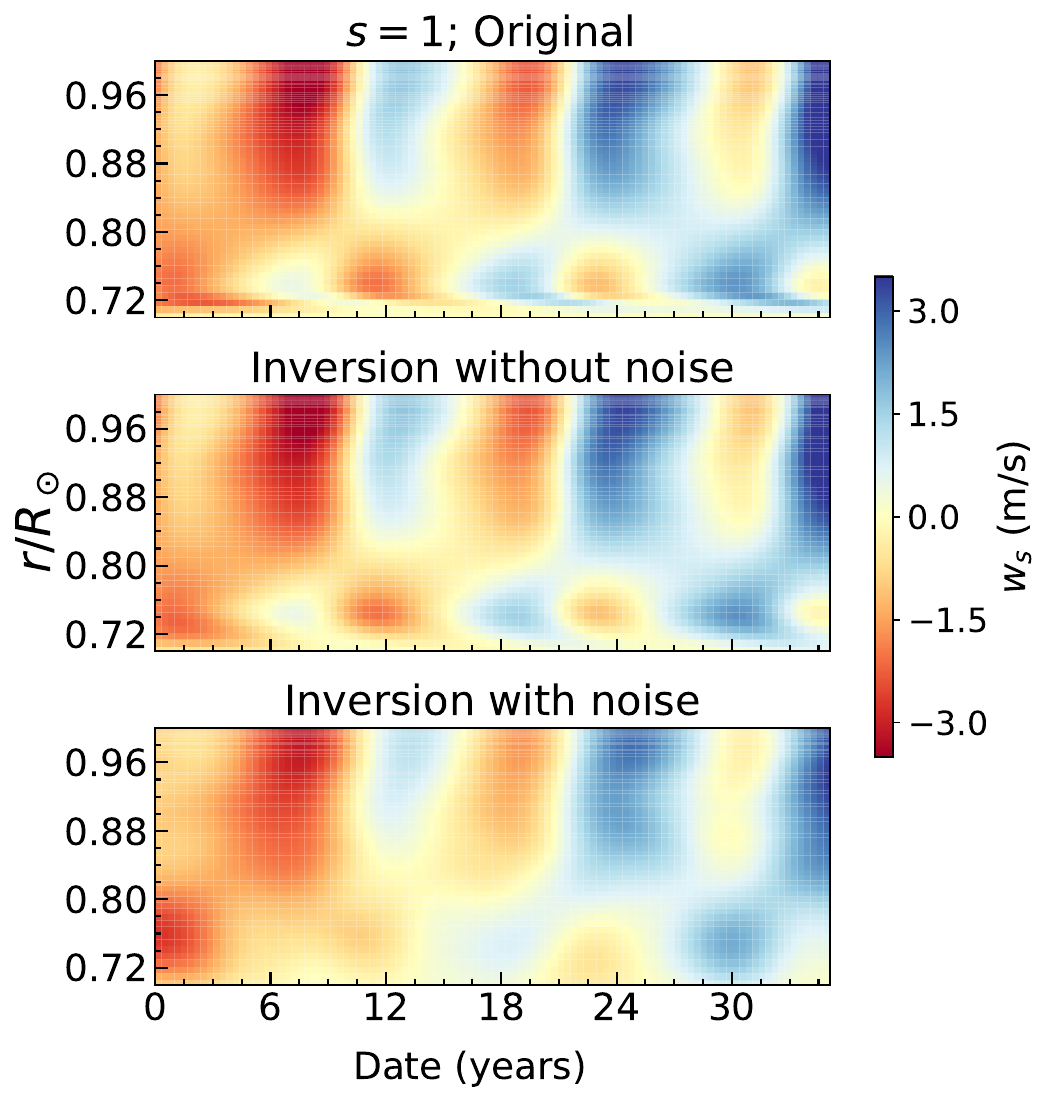}\includegraphics[scale=0.5]{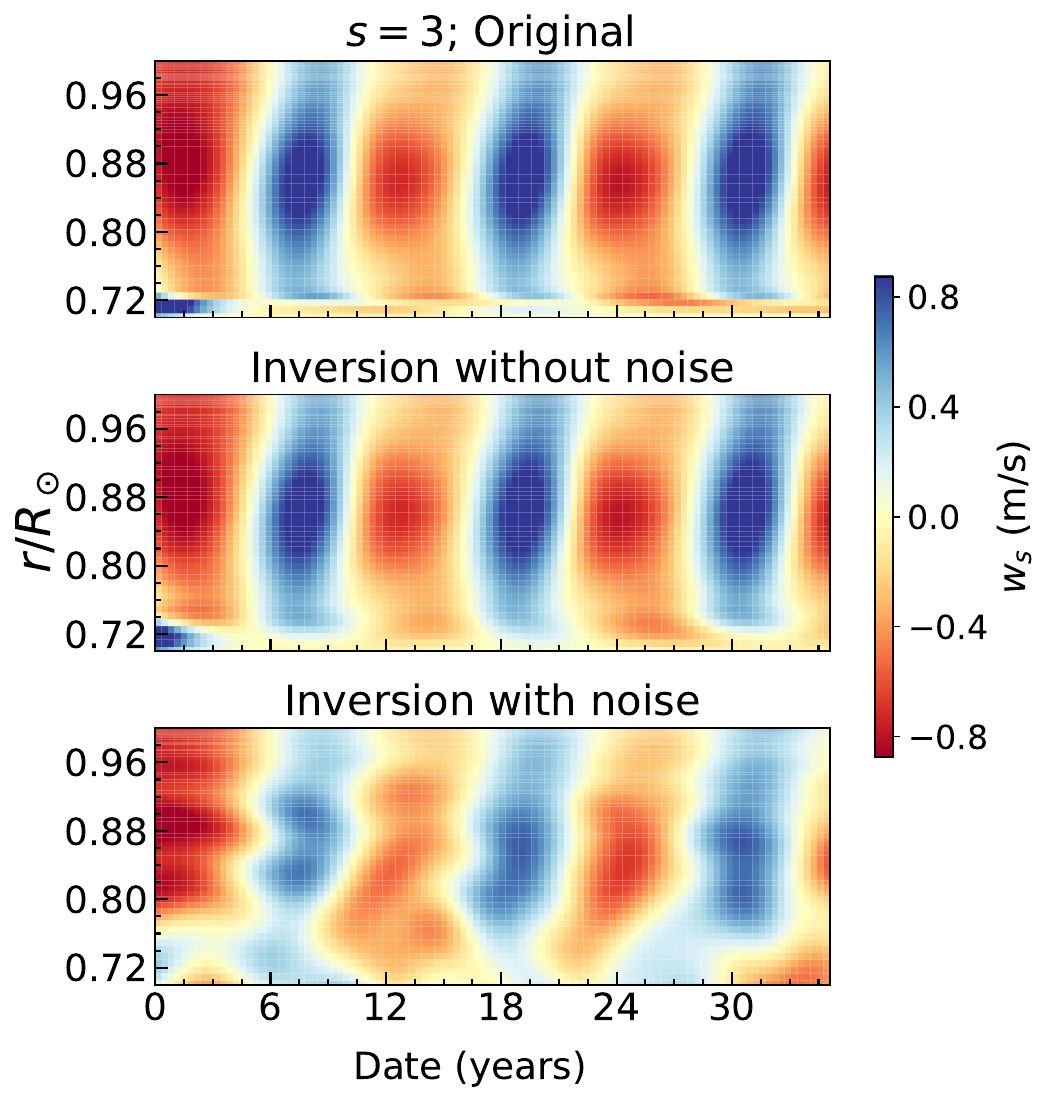}
     \caption{Zonal flow derived from the dynamo model undergoes spherical harmonic decomposition, as described by Equation \ref{eq:v_w}. The left and right panels illustrate our inversion outcomes for the $a$-coefficients of degrees $s=1$ and $s=3$, respectively. The top panel displays the original profile of $w_s$, utilized in forward modeling to derive the $a$-coefficients from Equation \ref{eq:a_coeff}. These $a$-coefficients serve as inputs for the inversion process, showcased in the middle panel. Following the initial calculation of the $a$-coefficients as previously mentioned, they are subsequently subjected to perturbation using Gaussian random noise proportional to the observed values. The inversion is then conducted based on this perturbed data, with the resulting analysis depicted in the bottom panel.}
     \label{fig:wst_inv}
 \end{figure}

\begin{figure}
 \centering
 \includegraphics[scale=0.5]{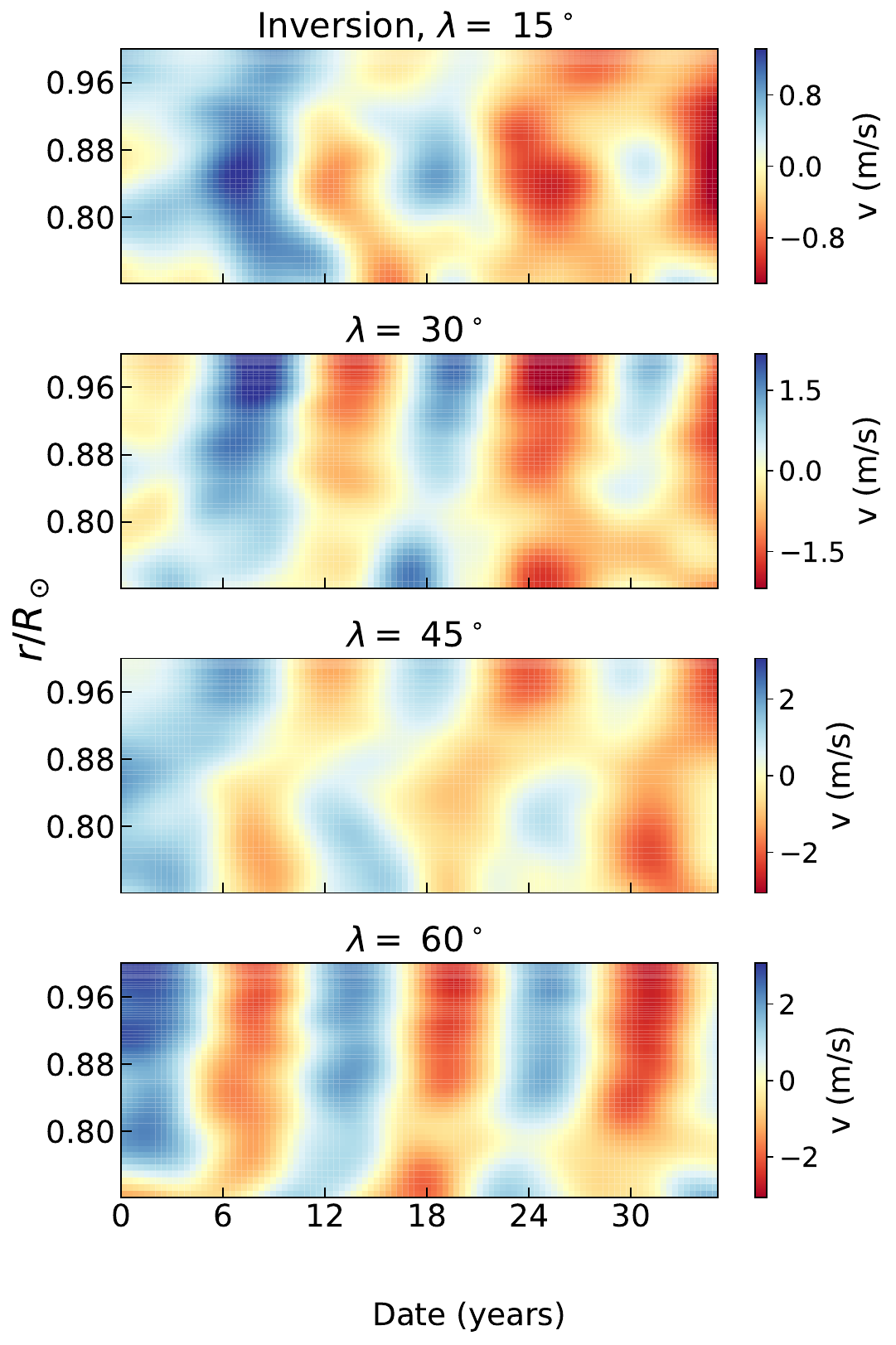}\includegraphics[scale=0.5]{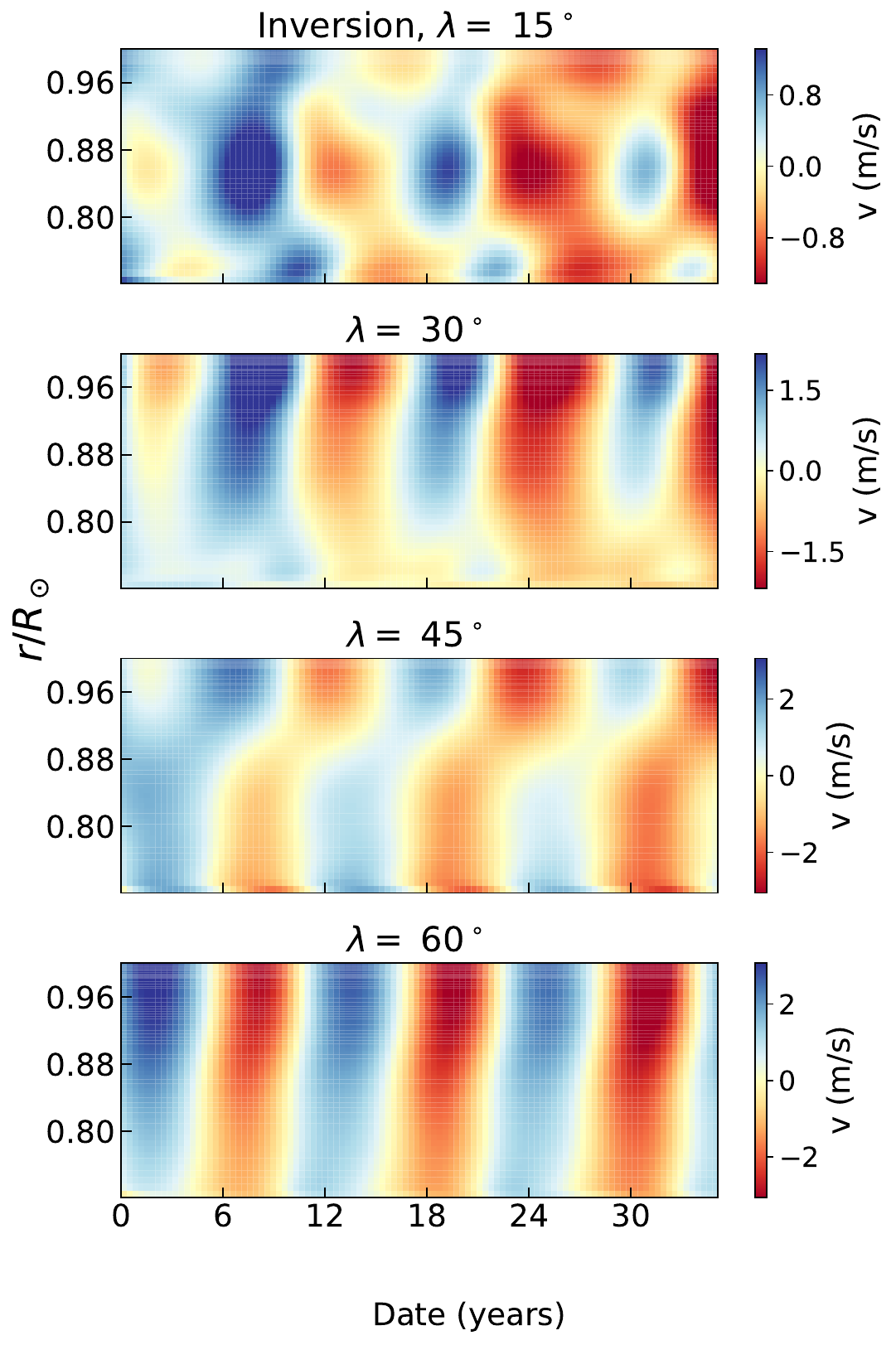}
    \caption{Zonal flow from the dynamo model is plotted against radius over time across various latitudes, indicated in each panel's title. The left panel illustrates the inverted profile, while the right panel displays the original profile. }
     \label{fig:zf_inv_r}
 \end{figure}

 We use noise from the $72$-day observational data to perturb the measurement. We use a model from \citet{pipin20} to generate artificial a-coefficients and perform inversion of these coefficients. In Figure \ref{fig:wst_inv}, we present a comparison between our inversion results and the original profile. We showcase results for both scenarios: without noise and with added noise. It's evident that without noise, the inversion performs exceptionally well. Even with added noise, the inversion effectively captures all the features of the dynamo wave pattern from the tachocline to the surface. We specifically show two harmonic degrees, $s=1$ and $s=3$.
  In Figure \ref{fig:zf_inv_r} within the dynamo model, we observe a band originating from the tachocline and migrating toward the surface. This migration takes longer near low-latitude regions ($<30^\circ$) but considerably less time at higher latitudes, around $60^\circ$. This identical profile has been decomposed into spherical harmonics to derive $w_s(r)$ using Equation \ref{eq:v_w}, facilitating forward modeling via Equation \ref{eq:a_coeff}. Having demonstrated in Figure \ref{fig:wst_inv} that we can accurately recover the original profile using our inversion methods, we combine components from all harmonic degrees to derive the velocity using Equation \ref{eq:v_w}. We compare our inverted profile against the original profile in Figure \ref{fig:zf_inv_r}.      
 \subsection{Accuracy near the tachocline}
\begin{figure}
 \centering
 \includegraphics[scale=0.7]{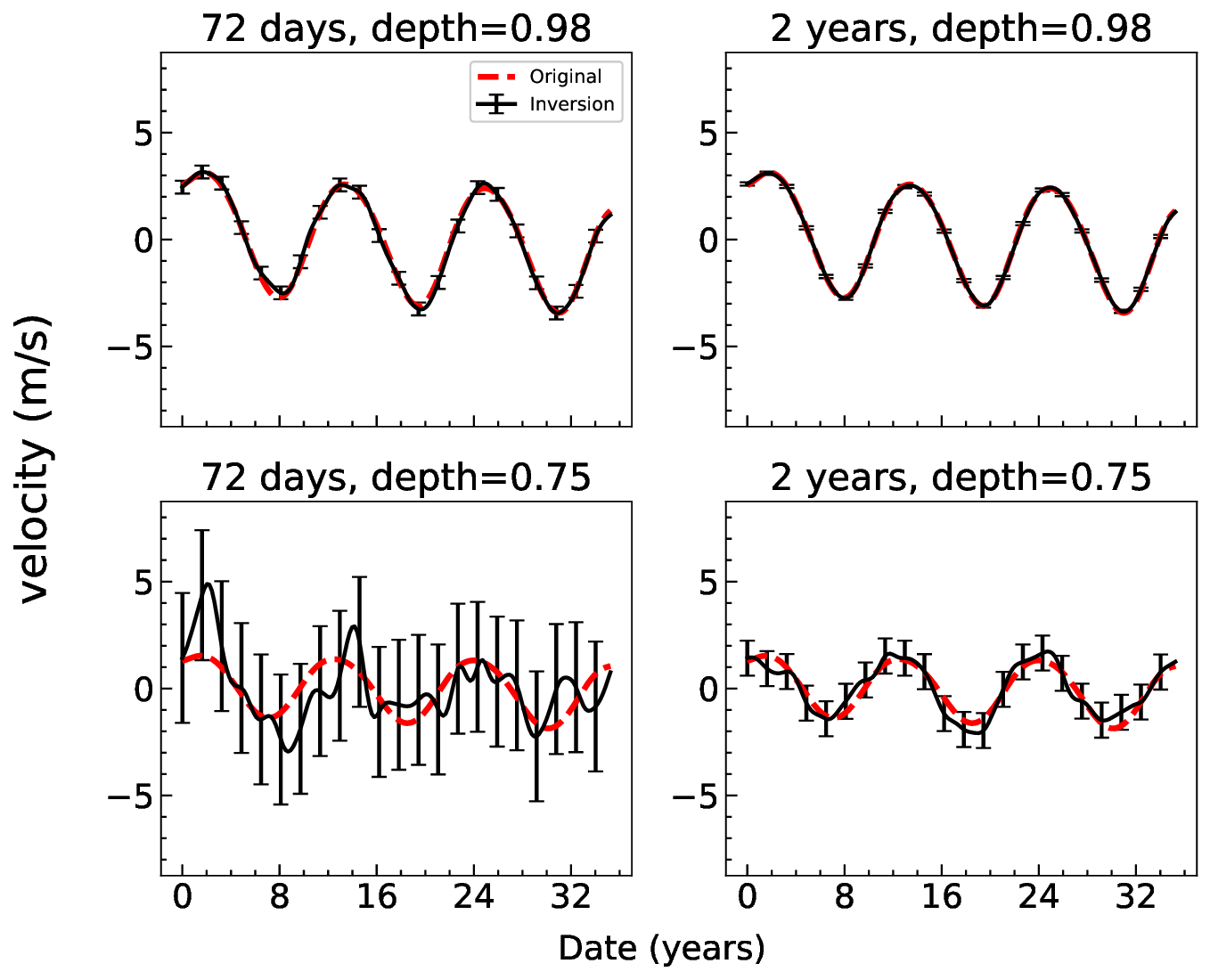}
    \caption{ We present our synthetic inversion results across varying noise levels. The left panels depict outcomes using noise akin to a $72$-day time series, while the right panel reflects results obtained from a $2$-year time series. All panels depict results corresponding to a latitude of $60^\circ$. Evidently, employing the $2$-year time series yields superior outcomes, particularly near the tachocline. However, both cases demonstrate equally commendable results near the surface at $0.98R_\odot$. To display the error bar at a depth of $0.98R_\odot$, we scaled it by a factor of $6$.}
     \label{fig:tacho}
 \end{figure}

\citet{sasha2019} discussed the potential for predicting the strength of future solar cycles by measuring torsional oscillation strength at the base of the convection zone, particularly at latitude $60^\circ$. To undertake a similar analysis, validating our accuracy near the base of the convection zone is imperative. To accomplish this, we assess the accuracy through synthetic tests. In Figure \ref{fig:tacho}, we compare our inverted values with the original ones at depth $0.75R_\odot$ and latitude $60^\circ$. We consider two cases. In one instance, we apply noise akin to a $72$-day measurement, while in another, we simulate noise comparable to a $2$-year measurement. We observe consistent accuracy in both scenarios near the surface layers, depth $0.98R_\odot$. However, there is a notable shift in the results as we approach the base of the convection zone. This variation aligns with expectations due to reduced sensitivity in deeper layers. P-modes spend less time traversing these depths, primarily because of the increasing sound speed from the surface towards the center of the Sun. This leads to an increase in the inversion error bars from the surface towards the center. As the error bars in splitting coefficients decrease with the square root of the total time, the inversion error bars near the tachocline tend to be smaller for longer time series measurements. This characteristic aids in interpreting variations in torsional oscillations around the tachocline. In Figure \ref{fig:tacho}, we illustrate this trend. JSOC typically provides the $a$-coefficient data for the time series spanning 72 days. However, \citet{sylvain23} has presented the splitting coefficients across various time series, including longer periods such as $1$ or $2$ years. This varied dataset might significantly reduce uncertainties near the base of the convection zone, aiding in understanding how torsional oscillation strength evolves over time. This understanding could be pivotal in forecasting the traits of forthcoming solar cycles \citep{sasha2019,pipin20}. However, it is important to note that long-term averaged data might blend various dynamical changes occurring at the tachocline. 

 \subsection{Inversion with observed data}  
\begin{figure}     
     \includegraphics[scale=0.5]{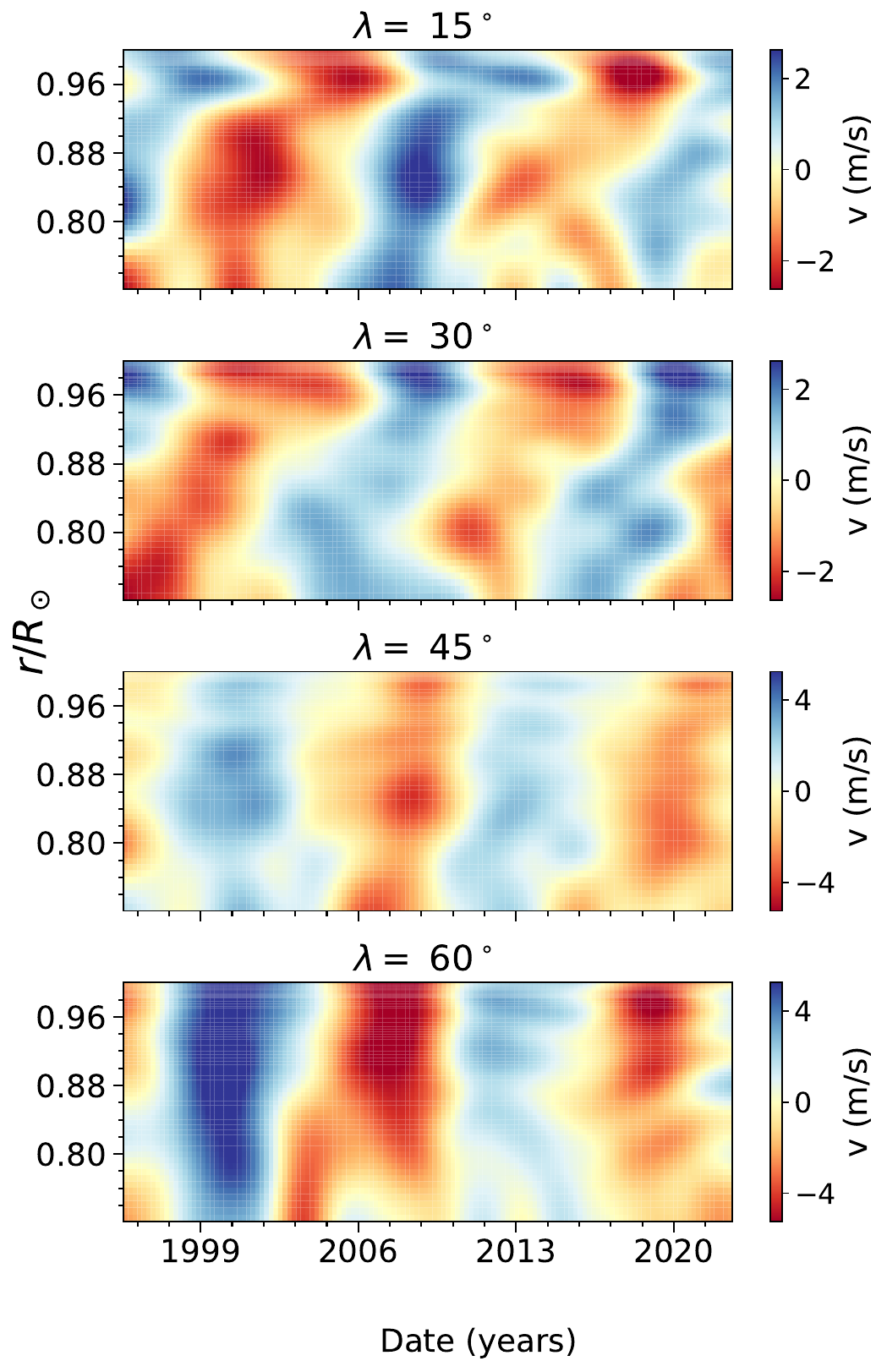}\includegraphics[scale=0.5]{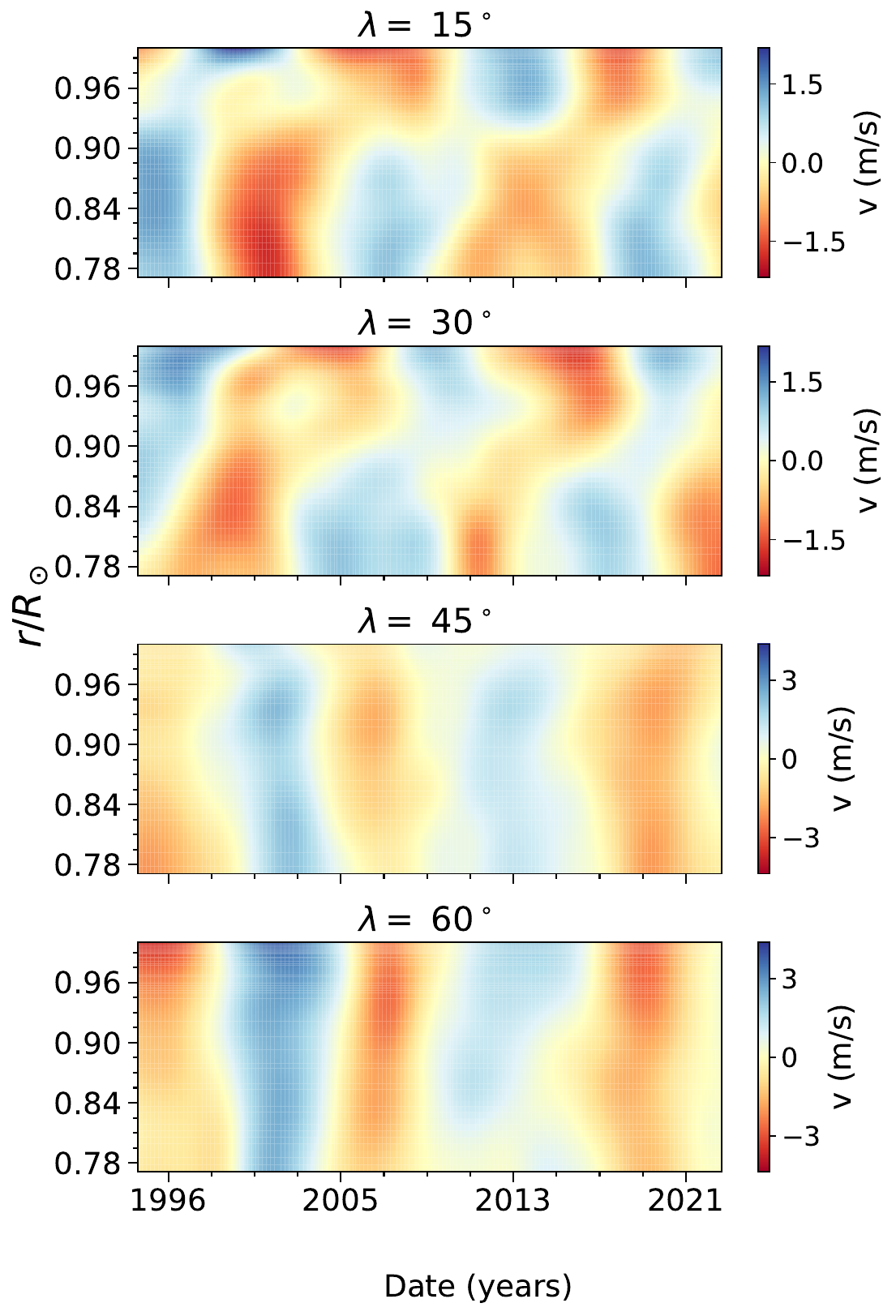}
     \caption{Zonal flow velocity is plotted as a function of time and radius for four latitudes, indicated in the title of each panel. The MDI and HMI results are shown in the left panel, and the GONG results are shown in the right panel.}
     \label{fig:zf_depth_year}
 \end{figure}

\begin{figure}     
     \includegraphics[scale=0.5]{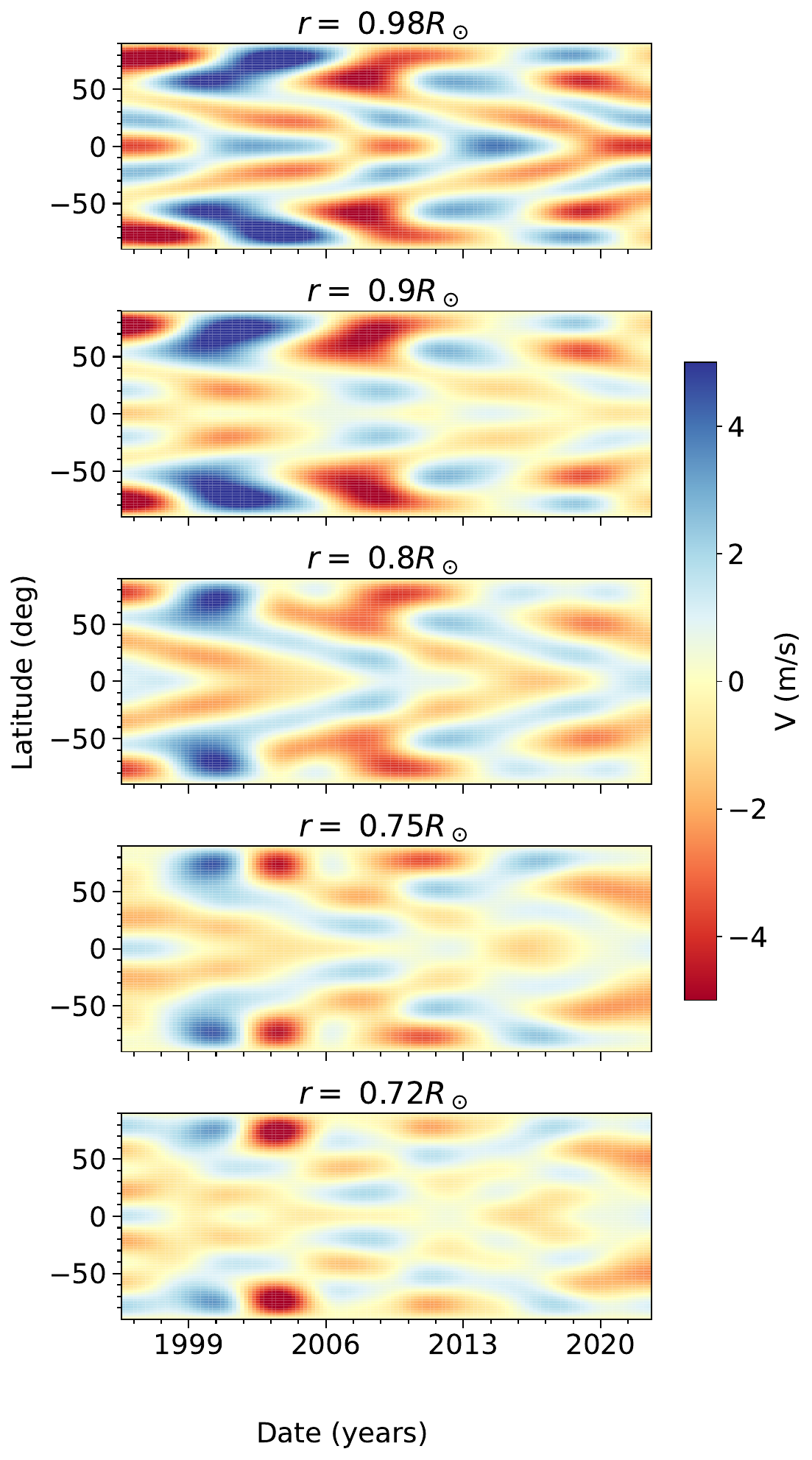}\includegraphics[scale=0.5]{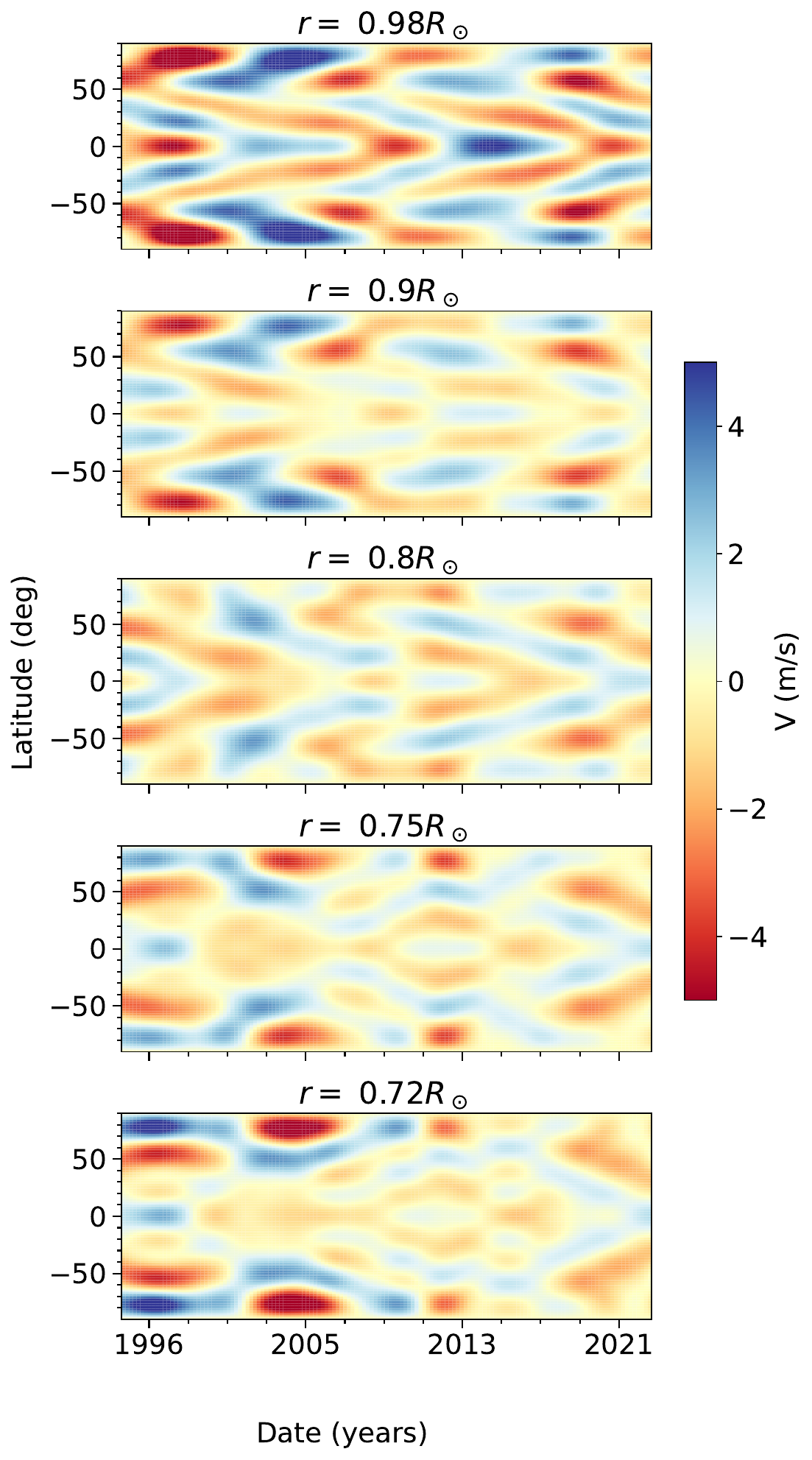}
     \caption{Zonal flow velocities plotted as a function of time and latitude for four depths mentioned in the title of each panel. The left panel illustrates the results derived from the combined MDI and HMI datasets, while the right panel showcases the findings specifically from the GONG datasets.}
     \label{fig:zf_latitude_year}
 \end{figure}

 After showcasing our ability to retrieve dynamo waves across the entire convection zone through the $72$-day measurement, we proceeded to conduct inversions using both MDI and HMI data sets, and during the overlap period, we used the HMI data. \citet{sasha2019} demonstrated the presence of dynamo waves in solar zonal flow by utilizing a-coefficients from the JSOC pipeline. To validate their analysis, we cross-checked using additional datasets from the NSO/GONG pipeline. Confirming the presence of the dynamo wave pattern in these alternative datasets would substantiate \citet{sasha2019} earlier detection. The results from the analysis of datasets collected by all three instruments are depicted in Figure \ref{fig:zf_depth_year} and \ref{fig:zf_latitude_year}. We employ a Gaussian filter with a one-year width in time to smooth the solution. In Figure \ref{fig:zf_latitude_year}, two branches of zonal flow are visible, both originating near mid-latitudes. One migrates towards the equator, while the other moves poleward. Zonal flow patterns persist across three solar cycles, encompassing cycles $23$, $24$, and ongoing cycle $25$. 
 From Figure \ref{fig:zf_depth_year}, the radial migration of the high-latitude branch is fast compared to the low-latitude branch, which takes a longer time to migrate from the tachocline to the surface.      


 \begin{figure}
     \centering
     \includegraphics[scale=0.5]{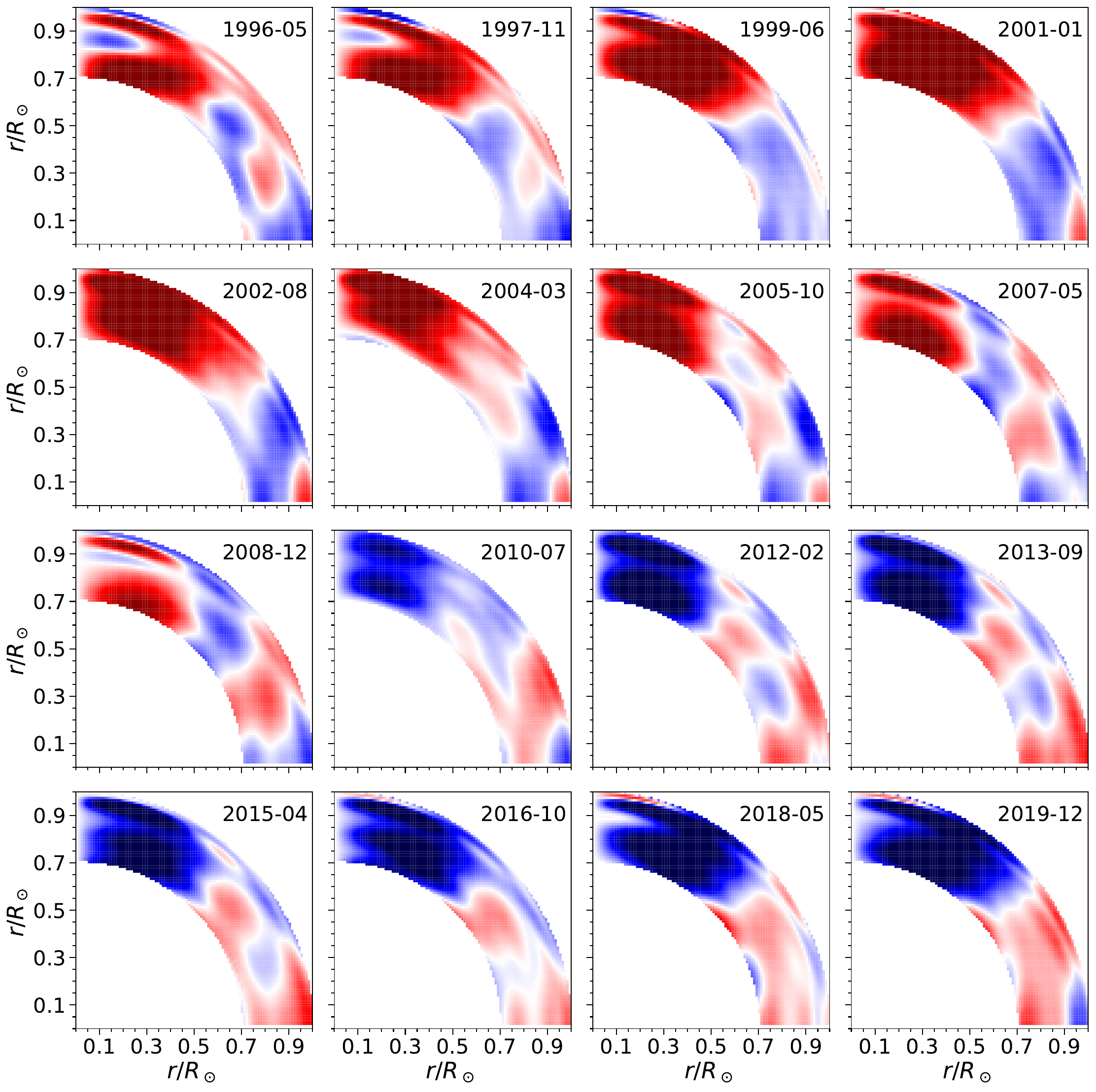}
     \caption{Evolution of solar zonal flow in the solar convection zone from the analysis of MDI and HMI combined data sets.}
     \label{fig:zf_r_theta}
 \end{figure}
 \section{Discussion and conclusions}
  The detection of dynamo waves holds significant importance in discerning the operational dynamo within the solar interior—whether it aligns with Babcock-Leighton or Parker's dynamo. This study aims to ascertain the feasibility of detecting dynamo-like wave patterns within the solar zonal flow. However, this detection poses a challenge due to its subtle nature compared to the prevailing mean rotation, which has a significantly higher magnitude. 
  
 Understanding the dynamo-wave-like structure within the zonal flows assumes crucial importance as it could elucidate the type of dynamo active in the Sun. This comprehension, in turn, might aid in predicting the characteristics of subsequent solar cycles, thereby contributing significantly to space weather prediction. Through forward modeling, we observe that variations in $a_s$ exhibit an $11$-year periodicity, paralleling similar characteristics observed in the a-coefficients depicted in Figure \ref{fig:a_coef_obs}, barring the potential for longer periodicity in $a_3$ coefficients. This forward modeling, along with the inversions, is necessary to decipher any subtle structure like dynamo waves present in the solar zonal flow measurements. After developing a forward modeling approach, we perform the inversion of the synthetic data sets. We show in Figure \ref{fig:zf_inv_r} that we can quite accurately recover the dynamo waves-like pattern in the zonal flow inferences from the tachocline to the surface. 
 
 After this testing, we used the observed data from three instruments MDI, HMI, and GONG, and found a similar dynamo-wave-like pattern in the inferred solar zonal flows. The similarity of the zonal flows deduced from the observations and the dynamo models allows us to link the observed variations to the dynamo waves, confirming the interpretation of \citet{sasha2019}.  We find that at high latitudes, the radial migration of the dynamo waves from the tachocline to the surface is very fast, whereas it is of the order of several years in the low-latitude areas. It exhibits a distinct behavior within the near-surface layers, where the zonal flows migrate from the surface to the bottom of the near-surface shear layer in the opposite direction compared to the rest of the convection zone, as illustrated in Figure \ref{fig:zf_depth_year}. This phenomenon mirrors a similar structure previously observed by \citet{vorontsov02} (refer to Figure $3$ in their study). Its origin has to be studied further. It can be connected  with the magnetic field  feedback on the on the angular momentum fluxes and the heat-transport  in the near-surface layers. 
 In the presented dynamo model, a similar feature can be noticed in the time-radius diagram at 45$^\circ$ latitude in Figure~\ref{fig:zf_inv_r}, but it is much less pronounced. 
 
 We can speculate a plausible dynamo scenario, previously discussed by \citet{sasha2019}, in which the generation of the poloidal magnetic field occurs due to helical turbulence in this high-latitude zone around $60^\circ$, near the base of the convection zone. The field swiftly migrates to the surface within a time frame of $1$ to $2$ years. However, in low latitudes, this migration process takes longer, reaching the surface after being twisted by differential rotation, transforming from the poloidal to the toroidal component. This aspect resembles Parker's dynamo theory \citep{parker_55_dynamo}, and is reproduced in our test dynamo model \citep{pipin2019}. Due to buoyancy instability, these magnetic fields emerge at the surface, forming sunspots. As sunspots decay, the remnant magnetic fields are transported by meridional flows to the polar regions, where they eventually descend and likely contribute to the seed field for the subsequent solar dynamo cycle. This facet of the dynamo process shares similarities with the models proposed by \citet{dikpati09} and \citet{Pipin2023}. Our results show clear signs of the  magnetic cycle  effects on  the near-surface dynamics of the torsional oscillations. 
 
Emerging dynamo models propose that the inflow toward active regions could influence the characteristics of the solar dynamo \citep{kinfe23}. This prompts an investigation into the potential implications of these inflows of the zonal flows and the corresponding observable in the frequency-splitting data. Both forward modeling and inversion techniques are essential to decipher the specific signature that should manifest in the frequency-splitting data, providing valuable insights into the solar dynamo's characteristics. 

In summary,  in this work, we developed a methodology for forward modeling of the effects of migrating zonal flows on the rotational frequency splitting data obtained from global helioseismology measurements and inversion for inferring the flows in the solar convection zone. Our inversion is based on the regularized least-squares RLS  method. We find that using our approach, we can detect the migrating dynamo-like wave patterns in the solar zonal flows using only 72-day measurements of p-modes of solar oscillations. However, the error bars are notably large near the tachocline, and our study shows that helioseismic analysis of the 2-year time series of solar oscillations  should be able to measure the zonal flows in the tachocline. We will address this task in future work. 

\par
\section*{}
\begin{flushleft}
\textit{Acknowledgments}: This work was partially supported by the NASA grants 80NSSC20K1320, 80NSSC20K0602, and 80NSSC22M0162. VP thanks the financial support of the Ministry of Science and Higher Education (Subsidy No.075-GZ/C3569/278).
\end{flushleft}
 
 \setlength{\parskip}{0pt}
 \appendix
 
\section{Theory} \label{sec:app_diffRot}
  
 \subsection{Properties of Generalized Spherical Harmonics}
 We use the following relations of generalized spherical harmonics 
\begin{align}
 Y_{\ell m}(\theta,\phi) & =\gamma_{\ell}Y_{\ell}^{0m}(\theta,\phi),\label{eq:sph_gen}\\
\frac{\partial Y_{\ell}^{Nm}}{\partial\theta} & =\frac{1}{\sqrt{2}}\left(\Omega_{N}^{\ell}Y_{\ell}^{N-1,m}-\Omega_{N+1}^{\ell}Y_{\ell}^{N+1,m}\right),\label{eq:dY_N}\\
-\frac{m}{\sin\theta}Y_{\ell}^{0m} & =\frac{\Omega_{0}^{\ell}}{\sqrt{2}}\left(Y_{\ell}^{-1,m}+Y_{\ell}^{1,m}\right)\\
\frac{N\cos\theta-m}{\sin\theta}Y_{\ell}^{Nm} & =\frac{1}{\sqrt{2}}\left(\Omega_{N}^{\ell}Y_{\ell}^{N-1,m}+\Omega_{N+1}^{\ell}Y_{\ell}^{N+1,m}\right)\\
\end{align}
where $\gamma_{\ell}=\sqrt{(2\ell+1)/(4\pi)}$ and $\Omega_{N}^{\ell}=\sqrt{\frac{1}{2}(\ell+N)(\ell-N+1)}$.
\subsection{Properties of Wigner-3j symbol}
We use the following symmetry properties of Wigner's 3j symbol
\begin{equation}
\begin{pmatrix}J_1 & J_2 & J_3\\
m_1 & m_2 & m_3
\end{pmatrix}=\begin{pmatrix}J_2 & J_3 & J_1 \\
m_2 & m_3 & m_1
\end{pmatrix}=\begin{pmatrix}J_3 & J_1 & J_2 \\
m_3 & m_1 & m_2
\end{pmatrix}
\label{eq:wigner_sym}
\end{equation}
For odd permutations we have 
\begin{equation}
    \begin{pmatrix}J_1 & J_2 & J_3\\
m_1 & m_2 & m_3
\end{pmatrix}=(-1)^{J_1+J_2+J_3}\begin{pmatrix}J_2 & J_1 & J_3\\
m_2 & m_1 & m_3
\end{pmatrix}
\end{equation}
The following recurrence relation will be necessary for our future analysis.
\begin{align}
\left[(j_{2}-m_{2})(j_{2}+m_{2}+1)(j_{3}+m_{3})(j_{3}-m_{3}+1)\right]^{1/2}\begin{pmatrix}j_{1} & j_{2} & j_{3}\\
m_{1} & m_{2}+1 & m_{3}-1
\end{pmatrix}\\
+\left[(j_{2}+m_{2})(j_{2}-m_{2}+1)(j_{3}-m_{3})(j_{3}+m_{3}+1)\right]^{1/2}\begin{pmatrix}j_{1} & j_{2} & j_{3}\\
m_{1} & m_{2}-1 & m_{3}+1
\end{pmatrix} & \nonumber\\
=\left(j_{1}(j_{1}+1)-j_{2}(j_{2}+1)-j_{3}(j_{3}+1)-2m_{2}m_{3}\right)\begin{pmatrix}j_{1} & j_{2} & j_{3}\\
m_{1} & m_{2} & m_{3}
\end{pmatrix}
\label{eq:wigner_rec}
\end{align}
 The integral of three generalized spherical harmonics obeys the following relations
\begin{align}
\int_{0}^{2\pi}d\phi\int_{0}^{\pi}d\theta\sin\theta\left(Y_{\ell^{\prime}}^{N^{\prime}m^{\prime}}\right)^{*}Y_{\ell^{\prime\prime}}^{N^{\prime\prime}m^{\prime\prime}}Y_{\ell}^{Nm}=4\pi(-1)^{(N^{\prime}-m^{\prime})}\begin{pmatrix}\ell^{\prime} & \ell^{\prime\prime} & \ell\\
-N^{\prime} & N^{\prime\prime} & N
\end{pmatrix}\begin{pmatrix}\ell^{\prime} & \ell^{\prime\prime} & \ell\\
-m^{\prime} & m^{\prime\prime} & m
\end{pmatrix}
\end{align}
It is to be noted that Wigner-3j symbols in Equation \ref{eq:wigner_sym} are non-zero when the following relation is satisfied
\begin{equation*}
    m1+m2+m3=0
\end{equation*}
Here we define another term 
\begin{equation}
    B^{(N)\pm}_{\ell^\prime \ell^{\prime\prime} \ell}=\frac{(-1)^N}{2}(1\pm (-1)^{\ell+\ell^\prime+\ell^{\prime\prime}})\left[\frac{(\ell^\prime+N)!(\ell+N)!}{(\ell-N)!(\ell-N)!}\right]\begin{pmatrix}\ell^{\prime} & \ell^{\prime\prime} & \ell\\
-N & 0 & N
\end{pmatrix}
\label{eq:B_N}
\end{equation}
One relation that shall be used to calculate the kernel is the following which can be derived from Equation \ref{eq:wigner_rec}
\begin{align}
    (\Omega^{\ell}_0)^2 B^{(1)-}_{s\ell\ell}+\Omega_{0}^{s}(\Omega^{\ell}_0)^2 \Omega^{\ell}_2(1-(-1)^{2\ell+s}\begin{pmatrix}\ell & s & \ell\\
1 & 1 & -2
\end{pmatrix}=\frac{1}{2}[2\ell(\ell+1)-s(s+1)]B^{(1)-}_{\ell s \ell},
    \label{eq:B_kernel}
\end{align}
\subsection{Kernel calculations}
Rotation frequency splitting is calculated using the perturbation theory based on the variational principle:
  \begin{equation}
\delta\omega_{k}=-i\frac{\int_{V}\rho_{0}\boldsymbol{\xi}_{k}^{*}\cdot(\mathbf{v}_{0}\cdot\nabla)\boldsymbol{\xi}_{k}dV}{\int_{V}\rho_{0}\vert\delta\mathbf{r}\vert^{2}dV},\label{eq:dOmega}
\end{equation}
$\boldsymbol{\xi}_{k}$ is the displacement eigenfunction of a mode, and $k$ signifies the combination of the radial order, angular degree, and azimuthal number  $(n,\ell,m)$ of a mode. The eigenfunction of a spherically symmetric Sun can be written as 
\begin{equation}
\boldsymbol{\xi}_{k}=U(r)Y_{\ell m}\hat{r}+V(r)\boldsymbol{\nabla}_{h}Y_{\ell m},
\label{eq:eigFn}
\end{equation}
  where $U(r)$ and $V(r)$ are the radial and horizontal components of the displacement vector. Reformulating Equation \ref{eq:eigFn} using generalized spherical harmonics and presenting them as components within a vector representation,
\begin{align}
    \xi_{k,r} & =\gamma_{\ell} U(r)Y_{\ell}^{0,m}\nonumber\\
    \xi_{k,\theta} & =\gamma_{\ell} V(r)\partial_{\theta} Y_{\ell}^{0,m}=\frac{\gamma_{\ell}}{\sqrt{2}}V(r)\Omega^{\ell}_0 (Y_{\ell}^{-1,m}-Y_{\ell}^{1,m})\nonumber\\
    \xi_{k,\phi} & = \gamma_{\ell} V(r)\partial_\phi Y_{\ell}^{0,m} = -i \frac{\gamma_{\ell}}{\sqrt{2}}V(r)\Omega^{\ell}_{0}(Y_{\ell}^{-1,m}+Y_{\ell}^{1,m})\\
    \label{eq:xi_sph}
\end{align}
The velocity of solar rotation can be written as
\begin{align}
\mathbf{v}_{0}(r,\theta) & =-\sum_{s=1,3,5}w_{s}^{0}(r)\partial_{\theta}Y_{s0}\hat{\phi}\nonumber\\
\mathbf{v}_{0}(r,\theta) & =-\sum_{s=1,3,5}\gamma_{s}w_{s}^{0}(r)\partial_{\theta}Y_{s}^{0,0}\hat{\phi}=-\sum_{s=1,3,5}\frac{\gamma_{s}}{\sqrt{2}}w_{s}^{0}(r)\Omega_{0}^{s}(Y_{s}^{-1,0}-Y_{s}^{1,0})\hat{\phi}
\label{eq:vel_sph}
\end{align}
We use Equation \ref{eq:dY_N} within Equation \ref{eq:vel_sph} to denote the derivative with respect to $\theta$. In order to calculate Equation \ref{eq:dOmega}, we need to define a tensor called $T$. 
\begin{equation}
T=\nabla\boldsymbol{\xi}_{k}
\end{equation}
The components of tensor $T$, which are required to calculate the sensitivity kernels are shown below. Temporarily, we omit the subscript $k$ from the eigenfunction $\boldsymbol{\xi}_k$ for the sake of brevity in notation and use the $r$, $\theta$, and $\phi$ components of the eigenfunction $\boldsymbol{\xi}_k$ from Equation \ref{eq:eigFn}.
\begin{align}
T_{\phi r} & =\frac{1}{r\sin\theta}(\partial_{\phi}\boldsymbol{\xi}).\hat{r}=\frac{1}{r\sin\theta}(\partial_{\phi}\xi_{r}-\xi_{\phi}\sin\theta)\nonumber\\
 & =\gamma_{\ell}U\frac{im}{r\sin\theta}Y_{\ell}^{0m}-(-i)\frac{\gamma_{\ell}}{\sqrt{2}}V\Omega_{0}^{\ell}r^{-1}\left(Y_{\ell}^{-1,m}+Y_{\ell}^{1,m}\right)\nonumber\\
 & =-ir^{-1}\gamma_{\ell}\frac{\Omega_{0}^{\ell}}{\sqrt{2}}(U-V)\left(Y_{\ell}^{-1,m}+Y_{\ell}^{1,m}\right),\label{eq:T_phi_r}\\
T_{\phi\theta} & =\frac{1}{r\sin\theta}(\partial_{\phi}\xi_{\theta}-\cos\theta\xi_{\phi})\nonumber\\
 & =r^{-1}\left(\frac{\gamma_{\ell}}{\sin\theta\sqrt{2}}V\Omega_{0}^{\ell}\partial_{\phi}(Y_{\ell}^{-1,m}-Y_{\ell}^{1,m})+\cot\theta i\frac{\gamma_{\ell}}{\sqrt{2}}V\Omega_{0}^{\ell}(Y_{\ell}^{-1,m}+Y_{\ell}^{1,m})\right)\nonumber\\
 & =ir^{-1}\frac{\gamma_{\ell}\Omega_{0}^{\ell}}{\sqrt{2}}V\left(\frac{m}{\sin\theta}(Y_{\ell}^{-1,m}-Y_{\ell}^{1,m})+\cot\theta(Y_{\ell}^{-1,m}+Y_{\ell}^{1,m})\right)\nonumber\\
 & =ir^{-1}\frac{\gamma_{\ell}}{2}V\Omega_{0}^{\ell}\Omega_{2}^{\ell}\left(Y_{\ell}^{2m}-Y_{\ell}^{-2m}\right),\label{eq:T_phi_theta}\\
T_{\phi\phi} & =\frac{1}{r\sin\theta}\left[\xi_{r}(\partial_{\phi}\hat{r})\cdot\hat{\phi}+\xi_{\theta}(\partial_{\phi}\hat{\theta})\cdot\hat{\phi}+\partial_{\phi}\xi_{\phi}\right] \nonumber\\
 & =\frac{1}{r\sin\theta}\left[\xi_{r}\sin\theta+\xi_{\theta}\cos\theta+\partial_{\phi}\xi_{\phi}\right] \nonumber\\
 & =\frac{1}{r}\left[\gamma_{\ell}U(r)Y_{\ell}^{0m}+\cot\theta\frac{\gamma_{\ell}}{\sqrt{2}}V(r)\Omega_{0}^{\ell}(Y_{\ell}^{-1,m}-Y_{\ell}^{1,m})+\frac{m}{\sin\theta}\frac{\gamma_{\ell}}{\sqrt{2}}V(r)\Omega_{0}^{\ell}(Y_{\ell}^{-1,m}+Y_{\ell}^{1,m})\right]\nonumber\\
 & =r^{-1}\left[\gamma_{\ell}U(r)Y_{\ell}^{0m}-\frac{\gamma_{\ell}}{2}V\Omega_{0}^{\ell}\left[\Omega_{2}^{\ell}\left(Y_{\ell}^{2m}+Y_{\ell}^{-2m}\right)+2\Omega_{0}^{\ell}Y_{\ell}^{0m}\right]\right]\nonumber\\
 & =r^{-1}\gamma_{\ell}\left[(U-\Omega_{0}^{\ell}\Omega_{0}^{\ell}V)Y_{\ell}^{0m}-V\Omega_{0}^{\ell}\Omega_{2}^{\ell}\frac{Y_{\ell}^{2m}+Y_{\ell}^{-2m}}{2}\right],,\label{eq:T_phi_phi}
\end{align}
 The numerator in Equation \ref{eq:dOmega} can be calculated using the tensor components of $T$ as following
\begin{equation}    
i\int\rho d\mathbf{r}\left[v_{\phi}(\xi_{k,r}^{*}T_{\phi r}+\xi^{*}_{k,\theta}T_{\phi\theta}+\xi^{*}_{k,\phi}T_{\phi\phi})\right]=T1+T2+T3,
\end{equation}
where 
\begin{align}
    T1=i \int \rho d\mathbf{r} v_{\phi}(\xi_{k,r}^{*}T_{\phi r}\nonumber\\
    T2=i \int \rho d\mathbf{r}v_{\phi} \xi^{*}_{k,\theta}T_{\phi\theta}\nonumber\\
    T3=i \int \rho d\mathbf{r}v_{\phi} \xi^{*}_{k,\phi}T_{\phi\phi}
    \label{eq:T_1_2_3}
\end{align}
We individually analyze each term in the above equation $T1$, $T2$, and $T3$ before combining them in the final step.
 \subsection{Analyzing $T1$ term}
 Substituting $v_\phi$, $\xi_{k,r}$ and $T_{\phi r}$ from Equations \ref{eq:vel_sph}, \ref{eq:xi_sph} and \ref{eq:T_phi_r} into Equation \ref{eq:T_1_2_3} for $T1$ term we get
\begin{align*}
i\int\rho r^2 dr\left(-\frac{\gamma_{s}}{\sqrt{2}}w_{s}^{0}\Omega_{0}^{s}(Y_{s}^{-1,0}-Y_{s}^{1,0})\right)(-ir^{-1})\gamma_{\ell}\frac{\Omega_{0}^{\ell}}{\sqrt{2}}(U-V)\left(Y_{\ell}^{-1,m}+Y_{\ell}^{1,m}\right)\gamma_{\ell}UY_{\ell}^{0m*}\\
=-4\pi\frac{\gamma_{\ell}^{2}\gamma_{s}\Omega_{0}^{\ell}\Omega_{0}^{s}}{2}\int\rho w_{s}^{0}d\mathbf{r}(U^{2}-UV)Y_{\ell}^{0m*}(Y_{s}^{-1,0}-Y_{s}^{1,0})\left(Y_{\ell}^{-1,m}+Y_{\ell}^{1,m}\right)\\
=-4\pi\frac{\gamma_{\ell}^{2}\gamma_{s}\Omega_{0}^{\ell}\Omega_{0}^{s}}{2}(-1)^{m}\begin{pmatrix}\ell & s & \ell\\
-m & 0 & m
\end{pmatrix}\left(1-(-1)^{2\ell+s}\right)\begin{pmatrix}\ell & s & \ell\\
0 & -1 & 1
\end{pmatrix}\int\rho w_{s}^{0} r^2dr(U^{2}-UV)\\
=-4\pi\frac{\gamma_{\ell}^{2}\gamma_{s}}{2}(-1)^{m}\begin{pmatrix}\ell & s & \ell\\
-m & 0 & m
\end{pmatrix}\int\rho w_{s}^{0}d\mathbf{r}(U^{2}-UV)B_{s\ell\ell}^{(1)-}
\end{align*}
In our case $m^{\prime}=m$ and selection rules of the 3j symbol
dictates
\begin{align*}
-N^{\prime}+N^{\prime\prime}+N & =0\\
-m^{\prime}+m^{\prime\prime}+m & =0=m^{\prime\prime}
\end{align*}
\subsection{Analyzing $T2$ term}
Substituting $v_\phi$, $\xi_{k,\theta}$ and $T_{\phi \theta}$ from Equations \ref{eq:vel_sph}, \ref{eq:xi_sph} and \ref{eq:T_phi_theta} into Equation \ref{eq:T_1_2_3} for $T2$ term we get
\begin{align*}
i\int\rho r^2 dr\left(-\frac{\gamma_{s}}{\sqrt{2}}w_{s}^{0}\Omega_{0}^{s}(Y_{s}^{-1,0}-Y_{s}^{1,0})\right)\left(ir^{-1}\frac{\gamma_{\ell}}{2}V\Omega_{0}^{\ell}\Omega_{2}^{\ell}\left(Y_{\ell}^{2m}-Y_{\ell}^{-2m}\right)\right)\left(\frac{\gamma_{\ell}}{\sqrt{2}}V(r)\Omega_{0}^{\ell}(Y_{\ell}^{-1,m}-Y_{\ell}^{1,m})\right)^{*}\\
=4\pi\frac{\gamma_{s}\gamma_{\ell}^{2}}{4}\Omega_{0}^{s}\left(\Omega_{0}^{\ell}\right)^{2}\Omega_{2}^{\ell}\int\rho rV^{2}w_{s}^{0}dr\begin{pmatrix}\ell & s & \ell\\
-m & 0 & m
\end{pmatrix}\left((-1)^{(-1-m)}\begin{pmatrix}\ell & s & \ell\\
1 & 1 & -2
\end{pmatrix}-(-1)^{1-m}\begin{pmatrix}\ell & s & \ell\\
-1 & -1 & 2
\end{pmatrix}\right)\\
=-4\pi\frac{\gamma_{s}\gamma_{\ell}^{2}}{4}\Omega_{0}^{s}\left(\Omega_{0}^{\ell}\right)^{2}\Omega_{2}^{\ell}(-1)^{m}\begin{pmatrix}\ell & s & \ell\\
-m & 0 & m
\end{pmatrix}\int\rho rV^{2}w_{s}^{0}dr(1-(-1)^{2\ell+s})\begin{pmatrix}\ell & s & \ell\\
1 & 1 & -2
\end{pmatrix}
\end{align*}
The term that we need to simplify is 
\begin{align*}
(-1)^{(-1-m)}\begin{pmatrix}\ell & s & \ell\\
1 & 1 & -2
\end{pmatrix}-(-1)^{1-m}\begin{pmatrix}\ell & s & \ell\\
-1 & -1 & 2
\end{pmatrix}\\
=-(-1)^{m}(1-(-1)^{2\ell+s})\begin{pmatrix}\ell & s & \ell\\
1 & 1 & -2
\end{pmatrix}
\end{align*}
where we have used the following identity 
\begin{equation}
\begin{pmatrix}J_{1} & J_{2} & J_{3}\\
m_{1} & m_{2} & m_{3}
\end{pmatrix}=(-1)^{J_{1}+J_{2}+J_{3}}\begin{pmatrix}J_{1} & J_{2} & J_{3}\\
-m_{1} & -m_{2} & -m_{3}
\end{pmatrix}
\end{equation}
\subsection{Analyzing $T3$ term}
Substituting $v_\phi$, $\xi_{k,\phi}$ and $T_{\phi \phi}$ from Equations \ref{eq:vel_sph}, \ref{eq:xi_sph} and \ref{eq:T_phi_phi} into Equation \ref{eq:T_1_2_3} for $T3$ term we get
\begin{align*}
T3 &=i\int\rho d\mathbf{r}\left(-\frac{\gamma_{s}}{\sqrt{2}}w_{s}^{0}\Omega_{0}^{s}(Y_{s}^{-1,0}-Y_{s}^{1,0})\right)\\
\left\{ r^{-1}\gamma_{\ell}
\left[(U-\Omega_{0}^{\ell}\Omega_{0}^{\ell}V)Y_{\ell}^{0m}-V\Omega_{0}^{\ell}\Omega_{2}^{\ell}\frac{Y_{\ell}^{2m}+Y_{\ell}^{-2m}}{2}\right]\right\}\\
\left(-i\frac{\gamma_{\ell}}{\sqrt{2}}V(r)\Omega_{0}^{\ell}(Y_{\ell}^{-1,m}+Y_{\ell}^{1,m})\right)^{*}\\
\end{align*}
We can simplify this equation as following
\begin{align}
T3=4\pi\frac{\gamma_{s}\gamma_{\ell}^{2}}{2}\Omega_{0}^{s}\Omega_{0}^{\ell}(-1)^{m}\int\rho r^{-1}w_{s}^{0}r^2 dr\begin{pmatrix}\ell & s & \ell\\
-m & 0 & m
\end{pmatrix}\nonumber\\
\left(-(1-(-1)^{2\ell+s})\begin{pmatrix}s & \ell & \ell\\
-1 & 0 & 1
\end{pmatrix}\left(UV-\Omega_{0}^{\ell}\Omega_{0}^{\ell}V^{2}\right)-\frac{V^{2}\Omega_{0}^{\ell}\Omega_{2}^{\ell}}{2}[1-(-1)^{2\ell+s}]\begin{pmatrix}\ell & s & \ell\\
1 & 1 & -2
\end{pmatrix}\right)&
\end{align}
After using the formula for $B^{(1)-}_{s\ell\ell}$ above Equation can be simplified further
\begin{align}
T3=4\pi\frac{\gamma_{s}\gamma_{\ell}^{2}}{2}(-1)^{m}\int\rho r^{-1}w_{s}^{0}d\mathbf{r}\begin{pmatrix}\ell & s & \ell\\
-m & 0 & m
\end{pmatrix}\nonumber\\
\left(\left(UV-\Omega_{0}^{\ell}\Omega_{0}^{\ell}V^{2}\right)B^{(1)-}_{s\ell\ell}-\frac{V^{2}\Omega_{0}^{s}(\Omega_{0}^{\ell})^2\Omega_{2}^{\ell}}{2}[1-(-1)^{2\ell+s}]\begin{pmatrix}\ell & s & \ell\\
1 & 1 & -2
\end{pmatrix}\right)
\end{align}
\subsection{Summing over all the terms}
After summing over $T1$, $T2$ and $T3$, coefficient associated with the term $V^2$ is following 
\begin{equation}
    -\Omega_{0}^{s}(\Omega^{\ell}_0)^2 \Omega^{\ell}_2 (1-(-1)^{2\ell+s})\begin{pmatrix}\ell & s & \ell\\
1 & 1 & -2
\end{pmatrix} - (\Omega^{\ell}_0)^2 B^{(1)-}_{s\ell\ell} =-\frac{1}{2}[2\ell(\ell+1)-s(s+1)]B^{(1)-}_{\ell s \ell}
\end{equation}
where we have used the identity Equation \ref{eq:B_kernel}. 
We get the equation for the kernel as follows
\begin{equation}
 M_{n\ell}= 4\pi\frac{\gamma_{s}\gamma_{\ell}^{2}}{2}(-1)^{m}\int\rho r^{-1}w_{s}^{0}d\mathbf{r} \begin{pmatrix}\ell & s & \ell\\
-m & 0 & m
\end{pmatrix}\left[ -U^2+ 2UV -\frac{V^2}{2}(2\ell(\ell+1)-s(s+1))\right]B^{(1)-}_{\ell s \ell},
\label{eq:kernel_exp}
\end{equation}
where the form for $B^{(1)-}_{\ell s \ell}$ from Equation \ref{eq:B_N} is the following 
\begin{equation}
    B^{(1)-}_{\ell s \ell}=\frac{-1}{2}(1-(-1)^{2\ell+s})\ell(\ell+1)\begin{pmatrix}\ell & s & \ell\\
-1 & 0 & 1
\end{pmatrix}
\end{equation}
 We can rearrange all the terms in the Equation \ref{eq:kernel_exp} and write it as follows
\begin{equation}
  M_{n\ell}= \sqrt{\frac{2s+1}{4\pi}}(2\ell+1)L^2 H_s^{1}H_s^{m}F_s^{2}\int\rho r^{-1}w_{s}^{0}d\mathbf{r} \left[ -U^2+ 2UV -\frac{V^2}{2}(2\ell(\ell+1)-s(s+1))\right]
  \label{eq:M_nl}
\end{equation} 
where we have used a similar notation as used by \citet{ritzwoller91} (Equation $32-35$ of their paper). 
\section{Asymptotic relation for the $a$-coefficients}
Matrix element due to frequency splitting is following (when $s<<l$)
\begin{align*}
M_{n\ell} & =-i\int_{0}^{R_\odot}\rho dV\bm{\xi}_{n\ell m}\cdot(\mathbf{u}\cdot\bm{\nabla}\bm{\xi}_{e\ell m})\\
 & =(-1)^{m+1}\left[\ell(\ell+1)(2\ell+1)\right]\sqrt{\frac{2s+1}{4\pi}}\sum_{s=1,2}\begin{pmatrix}s & \ell & \ell\\
0 & 1 & -1
\end{pmatrix}\begin{pmatrix}s & \ell & \ell\\
0 & m & -m
\end{pmatrix}\int_{0}^{R}\rho r^{2}dr\frac{w_{s}(r)}{r}\left[U^{2}-2UV+\frac{2\ell(\ell+1)-s(s+1)}{2}V^{2}\right]\\
 & =(-1)^{m+1}\left[\ell(\ell+1)(2\ell+1)\right]\sqrt{\frac{2s+1}{4\pi}}\sum_{s=1,2}(-1)^{\frac{s+2m+1}{2}}\frac{(s!!)^{2}}{2\ell^{2}s!}P_{s}(m/\ell)\langle\frac{w_{s}(r)}{r}\rangle_{n\ell}\\
 & \approx\sum_{s}\left((-1)^{(s-1)/2}\sqrt{\frac{2s+1}{4\pi}}\frac{(s!!)^{2}}{s!}\langle\frac{w_{s}(r)}{r}\rangle_{n\ell}\right)\ell P_{s}(m/\ell)\\
 & =\sum_{s}a_{s}\ell P_{s}(m/\ell)
\end{align*}
where the expression for $a_s$ is the following
\begin{equation}
    a_s(n,\ell)=(-1)^{(s-1)/2}\sqrt{\frac{2s+1}{4\pi}}\frac{(s!!)^{2}}{s!}\langle\frac{w_{s}(r)}{r}\rangle_{n\ell}
\end{equation}
where we have used the following simplified relation (Equation $A4$ from \citet{vorontsov11} )
\begin{equation}
\begin{pmatrix}s & \ell & \ell\\
0 & 1 & -1
\end{pmatrix}\begin{pmatrix}s & \ell & \ell\\
0 & m & -m
\end{pmatrix}=(-1)^{(s+2m+1)/2}\frac{(s!!)^2}{2\ell^2 s!}P_{s}(m/\ell) 
\end{equation}
\subsection{Full expression of kernel}
We can obtain relations for the $a$-coefficients as following from Equation \ref{eq:M_nl}
\begin{align}
    a_s=4\ell(\ell+1)(2\ell+1)\frac{2s+1}{4\pi}\frac{\{(2\ell-1)!\}^2}{(2\ell-s)!(2\ell+s+1)!}\mathcal{P}_s^{\ell}(1)\langle \frac{w_s(r)}{r}\rangle \nonumber\\
        =\int_{0}^{R}K_{s}(n,\ell;r)w_s(r)r dr  
    \label{eq:a_coeff_expr}
\end{align}
where the expression for $K_{s}(n,\ell)$ is following
\begin{equation}
    K_s(n,\ell;r)=4\ell(\ell+1)(2\ell+1)\frac{2s+1}{4\pi}\frac{\{(2\ell-1)!\}^2}{(2\ell-s)!(2\ell+s+1)!}\mathcal{P}_s^{\ell}(1)\rho\left[ U^2-2UV+\frac{2\ell(\ell+1)-s(s+1)}{2}V^2\right]
    \label{eq:kernel_expr}
\end{equation}

 \section{Inversion method}  \label{sec:app_inv}
 The frequency splitting is expressed in terms of the $a-$coefficients.
\begin{equation}
    \omega_{n\ell m}/(2\pi)= \nu_{n\ell}+\sum_{s=1,3,5,\dots,s_\text{max}} a_{n,\ell,s}\mathcal{P}_{s}^{\ell}(m)
\end{equation}
where splitting coefficients are related as follows
\begin{equation}
    a_{s}(n,\ell)=\int_{0}^{R_\odot} K_{n,\ell,s}(r)w_{s}(r)r^{2}dr,
\end{equation}
where $w_s$ are used to express velocity profile

\begin{equation}
    v(r,\theta)=\sum_{s}w_s(r)\partial_{\theta}Y_{s0}(\theta,\phi)
\end{equation}
We minimize the following equation to obtain $w_s(r)$ 
\begin{equation}
\chi=\sum_{n,\ell}\frac{1}{\sigma_{n,\ell}^{2}}\left(a_{n,\ell,s}-\int K_{n,\ell,s}(r)w_{s}(r)r^{2}dr\right)^{2}+\lambda\int\left(\frac{\partial^{2}w_{s}}{\partial r^{2}}\right)^{2}r^{2}dr,
\label{eq:chi_app}
\end{equation}
We express $w_{s}$ in terms of B-spline in radius as following
\begin{equation}
 w_{s}(r)=\sum_{i=1}^{N}b_{i}B_{i}(r),\label{eq:ws_Bspline}
\end{equation}
$N$ corresponds to the total number of unknowns. If we substitute Equation \ref{eq:ws_Bspline} into Equation \ref{eq:chi_app}, we get
\begin{align}
\chi=\sum_{n,\ell}\frac{1}{\sigma_{n,\ell}^{2}}\left(a_{s}(n,\ell)-\sum_{i}\int K_{n,\ell,s}(r)b_{i}B_{i}(r)r^{2}dr\right)^{2}+\lambda\int\left(\frac{\partial^{2}\sum_{i}b_{i}B_{i}(r)}{\partial r^{2}}\right)^{2}r^{2}dr
\end{align}
We minimize the above equation with respect to unknown coefficients $b_{j}$ to obtain 
\begin{align}
 \frac{\partial\chi}{\partial b_{j}}=-\sum_{n,\ell}\frac{1}{\sigma_{n,\ell}^{2}}\left(a_{s}(n,\ell)-\sum_{i}\int K_{n,\ell,s}(r)b_{i}B_{i}(r)r^{2}dr)\right)\int K_{n,\ell,s}(r)B_{j}(r)r^{2}dr \nonumber\\
 +\lambda\int\sum_{i}b_{i}\frac{\partial^{2}B_{i}(r)}{\partial r^{2}}r^{2}dr\int\frac{\partial^{2}B_{j}(r)}{\partial r^{2}}r^{2}dr
\end{align}
We define matrix $A$
\begin{align*}
A_{M,i} & =\int K_{M}B_{i}(r)r^{2}dr\\
D_{ij} & =\int\frac{\partial^{2}B_{i}(r)}{\partial r^{2}}r^{2}dr\int\frac{\partial^{2}B_{j}(r)}{\partial r^{2}}r^{2}dr
\end{align*}
where $M$ is a multi-index variable for $(n,\ell)$ and assume there are total $N_0$ such measurements. We get
\begin{equation}
\sum_{M\in N_0}\frac{1}{\sigma_{M}^{2}}A_{Mi}A_{Mj}b_{i}+\lambda\sum_{i}D_{ij}b_{i}=\sum_{M\in N_0}\frac{1}{\sigma_{M}^{2}}a_{M}A_{Mj}
\end{equation}
Which we can rewrite as 
\begin{equation}
\sum_{i=1}^{N}\left[\sum_{M\in N_0}(A^{T})_{jM}\Lambda_{MM}^{-1}A_{Mi}+\lambda D_{ij}\right]b_{i}=\sum_{M\in N_0}A_{jM}^{T}\Lambda_{MM}^{-1}a_{M},
\end{equation}
where $\Lambda$ is a diagonal matrix
\begin{equation}
\Lambda^{-1}=\begin{pmatrix}1/\sigma_{1}^{2} & 0 & 0 & 0& \dots\\
0 & 1/\sigma_{2}^{2} & 0 & 0 & \dots\\
0 & 0 & 1/\sigma_{3}^{2} & 0 & \dots\\
0 & 0 & 0 & 1/\sigma_{4}^{2} & \dots\\
\dots & \dots & \dots &\dots &\dots
\end{pmatrix}
\end{equation}
$\Lambda^{-1}$ is a $N_0\times N_0$ matrix. The above equation can be written in a matrix form as 
\begin{align}
\left[(A^{T}\Lambda^{-1}A)_{ji}+\lambda D_{ji}\right]b_{i} & =(A^{T}\Lambda^{-1})_{jM} a_{M}\nonumber\\
\left[A^{T}\Lambda^{-1}A+\lambda D\right]b & =A^{T}\Lambda^{-1}a
\label{eq:inv_app}
\end{align}
where matrix $b=(b_1,b_2,\dots,b_N)$. Solving these equations provides us with the values of $b$ which are used to reconstruct $w_{s}(r)$ from equation \ref{eq:ws_Bspline}.

\section{Averaging kernels} \label{sec:avg_kernel_app}
\begin{figure}
    \centering
     \includegraphics[scale=0.7]{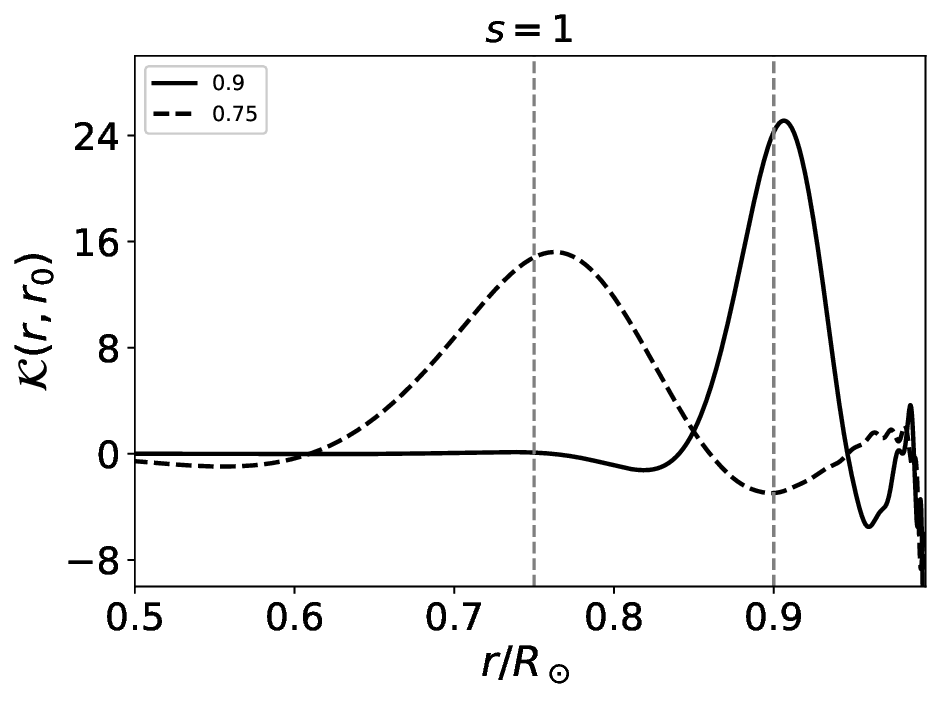}
     \caption{The averaging kernels for two distinct depths, $r=0.75R_\odot$ (indicated by a black dashed line) and $r=0.9R_\odot$ (shown as a black solid line). The target depths are highlighted by vertical dashed grey lines and are explicitly labeled in the legend of the plot. It is evident that the averaging kernel tends to have a narrower width closer to the surface, while it becomes broader in deeper layers.}
     \label{fig:avg_kernel}
\end{figure}

The splitting coefficients, $a_s$ are related to the velocity coefficients, $w_s$:  
\begin{equation}   
    a_{s}(n,\ell)=\int_{0}^{R_\odot} K_{s}(n,\ell;r)w_{s}(r)r^2 dr,
     \label{eq:a_coeff_App}  
\end{equation}
We look for a linear combination of $a_s(n,\ell)$ such that we can get an estimate of $w_s(r)$ at a target depth $r_0$ as following 
\begin{equation}
    \bar{w}_s(r)=\sum_{M \in N_0} C_M a_s(M),
    \label{eq:avg_kernel_c}
\end{equation}
$C_{M}$ are the coefficients that need to be determined. Therefore, formula for the averaging kernels is 
\begin{equation}
    \mathcal{K}(r,r_0)=\sum_{M}C_M K_s(M;r),
    \label{eq:avg_kernel_app}
\end{equation}
From Section \ref{sec:app_inv} we know 
\begin{equation}
    w_s(r_0)=\sum_{i=1}^N b_i B_{i}(r_0)        
\end{equation}
where $b_i$ is derived by performing the matrix operations outlined in Equation \ref{eq:inv_app}.
\begin{equation}
 b=(A^T\Lambda^{-1} A+\lambda D)^{-1} (A^T \Lambda^{-1})a
 \label{eq:}
\end{equation}
Since $b$ and $w_s$ are related through $w_s(r)=\sum_i b_i B_i(r)$, we can find the equation for coefficients $c_M$ using the following equations
\begin{equation}
    C_M=\sum_{j} B_j(r_0) ((A^T\Lambda^{-1} A+\lambda D)^{-1} (A^T \Lambda^{-1}))_{jM}
\end{equation}
or in matrix notation
\begin{equation}
    C=\mathbf{B}  ((A^T\Lambda^{-1} A+\lambda D)^{-1} (A^T \Lambda^{-1})),
\end{equation}
where $B=[B_1(r_0),B_2(r_0), \cdots]$ represents a matrix created by applying B-spline basis functions evaluated at $r_0$. After determining the coefficients $C$, we substitute them into Equation \ref{eq:avg_kernel_app} to derive the averaging kernel at the target depth $r_0$. In Figure \ref{fig:avg_kernel}, we display averaging kernel at two different depths $0.75R_\odot$ and $0.9R_\odot$. 
\subsection{Averaging Kernels in 2D}
\begin{figure}   
 \centering
    \includegraphics[scale=0.6]{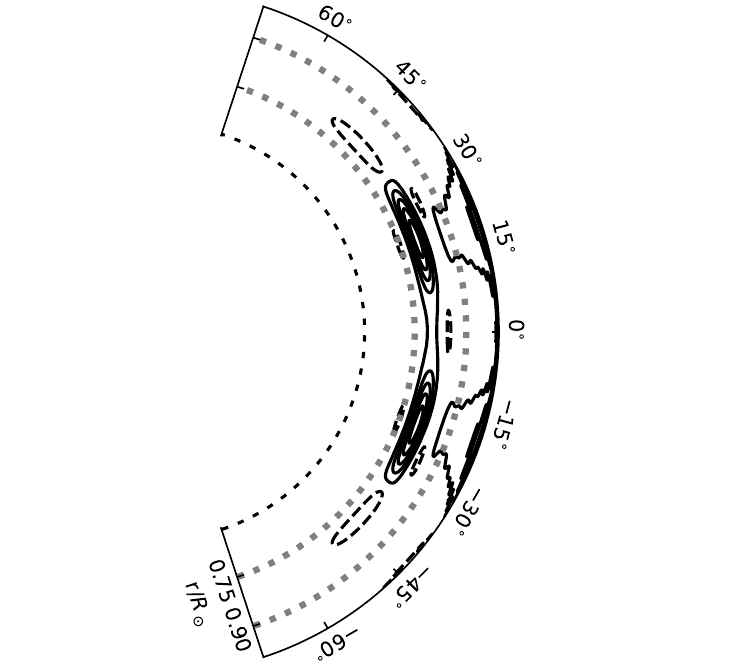}\includegraphics[scale=0.6]{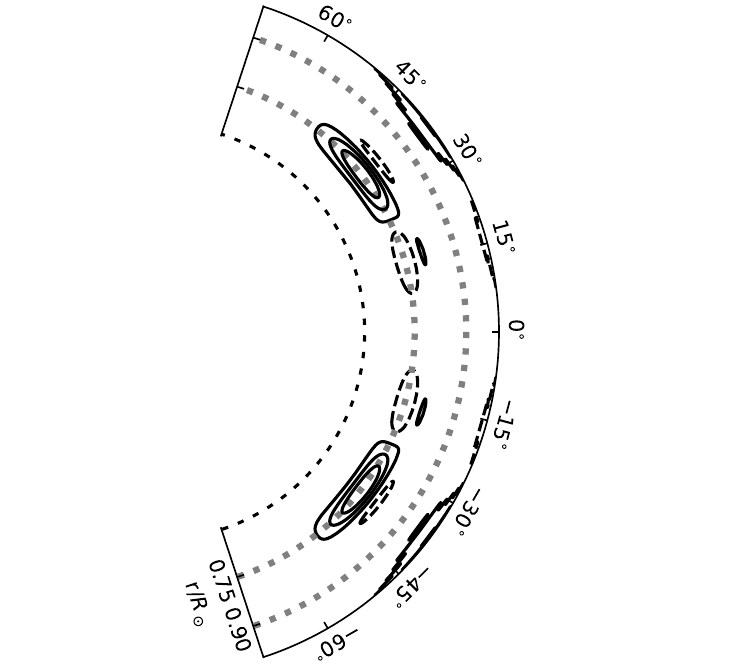}
    \caption{Left panel: averaging kernel plotted at radius $0.8R_\odot$ and latitude $20^\circ$. Right panel: averaging kernel plotted at radius $0.75R_\odot$ and latitude $40^\circ$.}
    \label{fig:avg_kernel_2D}
\end{figure}

Solar rotation, $\Omega(r,\theta)$ can be expressed as a toroidal function 
\begin{equation}
    \Omega(r,\theta)=\sum_{s} w_s(r)\Psi_s(\theta)
    \label{eq:omega_app}
\end{equation}
 where $w_s$ is obtained from Equation \ref{eq:avg_kernel_c} as a linear combination of splitting coefficients $a_s$ as following
 \begin{equation}
    w_s(r)=\sum_{n,\ell}C_s(r;n,\ell) a_s(n,\ell)    
    \label{eq:w_s_app}
 \end{equation}
We can find a relationship between the $a-$coefficients and frequency splitting by minimizing the following misfit function \citep{pijpers97}
\begin{equation}
    \chi^2=\sum_{m=-\ell}^{\ell}\left( \nu_{n\ell m}-\sum_{s=0}^{s_{\text{max}}}a_{s}(n,\ell) \mathcal{P}_{s}^{\ell}(m)\right) ^2
\end{equation}
 with respect to $a_s$. This yields
 \begin{equation}
     a_{s}(n,\ell)=\frac{\sum_{-\ell}^{\ell}\nu_{n\ell m}\mathcal{P}_{s}^{\ell}(m)}{ \sum_{m=-\ell}^{\ell}\mathcal{P}_{s}^{\ell}(m)^2}
     =\sum_{m=-\ell}^{\ell}A_{s}(n,\ell,m)\nu_{n\ell m} 
    \label{eq:a_s_app}
 \end{equation}
 where $A_{s}(n,\ell,m$ is defined as 
 \begin{equation}
     A_s(n,\ell,m)=\frac{\mathcal{P}_{s}^{\ell}(m)}{\sum_{m^\prime=-\ell}^{\ell}\left(\mathcal{P}_{s}^{\ell}(m^\prime)\right)^2}
 \end{equation}
 Substituting Equation \ref{eq:a_s_app} and \ref{eq:w_s_app} into Equation \ref{eq:omega_app} we get
 \begin{align}
     \Omega(r,\theta)=\sum_{s}\sum_{n,\ell}C_s(r;n,\ell)\left(\sum_{m=-\ell}^{\ell}A_{s}(n,\ell)\nu_{n\ell m} \right)\psi(\theta)\nonumber\\
     =\sum_{n\ell m} C_s(r;n,\ell)A_{s}(n,\ell)\psi(\theta)\nonumber\\
     =\sum_{n\ell m}C_{n\ell m}^{2D}\nu_{n\ell m},
    \label{eq:omega_linear}
 \end{align}
 where
 \begin{equation}
     C_{n\ell m}^{2D}=\sum_{s}c_{s}(r;n,\ell)A_{s}(n,\ell)\psi(\theta),
     \label{eq:c_coeff}
 \end{equation}
 Therefore, we can write the equation for the averaging kernels as the following
 \begin{equation}
     \mathcal{K}(r,\theta,r_0,\theta_{0})=\sum_{n\ell m}C_{n\ell m}^{2D}(r_0,\theta_0) K_{n\ell m}(r,\theta),
     \label{eq:avg_kernel_2D}
 \end{equation}
 where $K_{n\ell m}(r,\theta)$ is the $2-$D kernel defined as 
 \begin{equation}
    K_{n\ell m}(r,\theta)= \frac{R_{n\ell m}}{I_{n\ell m}}
 \end{equation}
 where $R_{n\ell m}$ is defined as the following
 \begin{align}
    R_{n\ell m} &= m \rho r \sin \theta \Bigg\{ \lvert U \rvert^2 P_{\ell}^{m}(\cos\theta)^2 + \lvert V(r) \rvert^2 \Bigg[ \left( \frac{dP_{\ell}^{m}}{d\theta} \right)^2 \nonumber\\
    &\quad + \frac{m^2}{\sin^2\theta}P_{\ell}^{m}(\cos\theta)^2 \Bigg] - P_{\ell}^{m}(\cos\theta)^2[U^{*}V+U V^{*}] \nonumber\\
    &\quad - 2P_{\ell}^{m}(\cos\theta)\frac{dP_{\ell}^{m}}{d\theta}\frac{\cos\theta}{\sin\theta}\lvert V \rvert^2 \Bigg\} 
\end{align}
 where $I_{n\ell m}$ is 
 \begin{equation}
    I_{n\ell m}= \frac{2}{2\ell+1}\frac{(\ell+\vert m \vert)!}{(\ell-\vert m\vert)!}\int_{0}^R \left[U^2 +\ell(\ell+1)V^2\right]\rho r^2
dr \end{equation}
In Figure \ref{fig:avg_kernel_2D}, we present the averaging kernels for various depths and latitudes. The figure clearly illustrates a distinctive localization in both latitude and radius.

\bibliography{references} 
  
\end{document}